\documentclass[11pt,a4paper]{article}
\pdfoutput=1
\usepackage{jstyle}
\usepackage{amsmath, amsfonts, amssymb}
\usepackage{graphicx, hyperref, verbatim}
\usepackage[dvipsnames]{xcolor}
\usepackage{float}
\usepackage{caption}
\usepackage{subcaption}

\DeclareGraphicsExtensions{.pdf,.png,.jpg,.mps}
\usepackage{booktabs}
\usepackage{multirow}
\usepackage{tikz}

\usetikzlibrary{calc,matrix,decorations.pathmorphing,decorations.markings,arrows,positioning,intersections,mindmap,backgrounds,patterns,arrows.meta,chains}
\usepackage{cancel}

\usepackage{physics}
\usepackage{diagbox}
\usepackage{pifont}
\usepackage{cleveref}

\newcommand{\starlight}{{\text{\ding{71}}_{\text{\hspace{-1.3pt}\ding{71}}}}}
\newcommand{\doubledagger}{{{\dagger}_{\hspace{-1.3pt}\dagger}}}
\newcommand{\doubleflake}{{\text{\ding{93}}_{\text{\hspace{-1.3pt}\ding{93}}}}}

\newcommand{\zb}{{\bar{z}}}

\newcommand{\dst}{{\mathtt{d}_{st}}}
\newcommand{\dsu}{{\mathtt{d}_{su}}}
\newcommand{\dtu}{{\mathtt{d}_{tu}}}

\newcommand{\tmp}{\omega}
\newcommand{\DU}{\mathcal{D}_{U}}
\newcommand{\DV}{\mathcal{D}_{V}}

\title{{A differential representation for\\ holographic correlators}}

\author[]{Zhongjie Huang$^{\starlight}$,}
\author[]{Bo Wang$^{\doubledagger}$,}
\author[]{Ellis Ye Yuan$^{\doubleflake}$,}
\affiliation[]{Zhejiang Institute of Modern Physics, School of Physics, Zhejiang University, \\Hangzhou, Zhejiang 310058, China }
\affiliation[]{Joint Center for Quanta-to-Cosmos Physics, Zhejiang University,
	\\Hangzhou, Zhejiang 310058, China}

\abstract{We present a differential representation for holographic four-point correlators. In this representation, the correlators are given by acting differential operators on certain seed functions. The number of these functions is much smaller than what is normally seen in known examples of holographic correlators, and all of them have simple Mellin amplitudes. This representation establishes a direct connection between correlators in position space and their Mellin space counterpart. The existence of this representation also imposes non-trivial constraints on the structure of holographic correlators. We illustrate these ideas by correlators in ${\rm AdS}_5 \times {\rm S}^5$ and ${\rm AdS}_5 \times {\rm S}^3$.}

\emailAdd{$^{\starlight}$}{zjhuang@zju.edu.cn}
\emailAdd{$^{\doubledagger}$}{b\_w@zju.edu.cn}
\emailAdd{$^{\doubleflake}$}{eyyuan@zju.edu.cn}

\begin{document} 
\maketitle
\tableofcontents
	
\newpage

\section{Introduction}\label{sec:Introduction}

In recent years there has been a rapid growth of analytic results for four-point correlators of protected operators in various holographic models, which far surpass the applicability of traditional Witten diagrams. For example, at four points this allows to analytically determine tree-level correlators of all half-BPS operators in type IIB supergravity on $\mathrm{AdS}_5\times\mathrm{S}^5$ \cite{Rastelli:2016nze,Rastelli:2017udc}, and further construct various correlators at one and two loops \cite{Alday:2017xua,Aprile:2017bgs,Alday:2017vkk,Alday:2019nin,Aprile:2017qoy,Aprile:2019rep,Huang:2021xws,Drummond:2022dxw}. This success is achieved by the novel idea of merging the bulk intuition of perturbative expansion and estimation of operator spectrum into the analytic bootstrap methods of boundary CFTs. See \cite{Bissi:2022mrs,Heslop:2022xgp} for detailed reviews on this method as well as its broader applications.

An interesting theme in this area of exploration is the choice of representations for the conformal correlators. They can be either expressed as functions of position space coordinates, or via Mellin transformations, equivalently expressed as Mellin amplitudes \cite{Mack:2009mi}. While the former has close contact with position-space physics, the latter brings the correlator into a form analogous to flat-space scattering amplitudes \cite{Penedones:2010ue,Fitzpatrick:2011ia}. 

To be concrete, take the prototypical model of type IIB supergravity in $\mathrm{AdS}_5\times\mathrm{S}^5$ for instance. Here the bulk tree-level contribution to the four-point correlator $\langle\mathcal{O}_2\mathcal{O}_2\mathcal{O}_2\mathcal{O}_2\rangle$ of the simplest half-BPS operator $\mathcal{O}_2$ can be reduced to a function of two conformal cross ratios $\mathcal{H}_{2222}(U,V)$. This function is related to a Mellin amplitude of two Mellin-Mandelstam variables $\mathcal{M}_{2222}(s,t)$ via the inverse Mellin integral
\begin{equation}\label{eq:example2222Mt}
	\mathcal{H}_{2222}(U,V) = \int \frac{\dd s \dd t}{(2\pi i)^2}\, U^{\frac{s+4}{2}}V^{\frac{t-4}{2}}\mathcal{M}_{2222}(s,t)\, \Gamma^2(2-\frac{s}{2})\Gamma^2(2-\frac{t}{2})\Gamma^2(\frac{s+t}{2})\,,
\end{equation}
On the one hand, the position space result for this correlator can be written as a single scalar contact Witten diagram (which is denoted by the so-called $\bar{D}$-functions) \cite{Arutyunov:2000py}
\begin{equation}\label{eq:example2222p}
	\begin{split}
		\mathcal{H}_{2222}(U,V) =&-U^4 \bar{D}_{2422}(U,V)\\
		= &\quad \frac{P_0(z,\zb)}{(z-\bar{z})^4} + \frac{P_{1,1}(z,\zb)}{(z-\bar{z})^6}\log U + \frac{P_{1,2}(z,\zb)}{(z-\bar{z})^6}\log V\\
		&+ \frac{P_2(z,\zb)}{(z-\bar{z})^7} \left[2{\rm Li}_2(z)-2{\rm Li}_2(\zb) + \log(z\zb)\log\left(\frac{1-z}{1-\zb}\right)\right]\,, 
	\end{split}
\end{equation}
where $P(z,\bar{z})$ are some polynomials of $z$ and $\bar{z}$, whose explicit expressions are not important for the present. Information on operator spectrum and the OPE coefficients is directly enoded in the conformal block expansion of this function. On the other hand, the Mellin representation of the same correlator is given by \cite{Rastelli:2016nze,Rastelli:2017udc}
\begin{equation}\label{eq:example2222M}
	\mathcal{M}_{2222}(s,t) = \frac{2}{(s-2)(t-2)(2-s-t)}\,. 
\end{equation}
This is structurally almost the same as the graviton four-point scattering amplitude in 10d IIB supergravity. As can be expected, this representation greatly simplifies the study of flat-space limit (or large AdS radius limit) of the correlator \cite{Penedones:2010ue}. 

Despite of the established transformation \eqref{eq:example2222Mt} between the two representations, it is very often a non-trivial task to fully verify their equivalence. As a result, properties and patterns that are natural and simple on one side involve huge efforts to understand on the other side. For example, in the above case we observe a correspondence between a rational function in Mellin space and polylogarithms in the position space. However, it is far from obvious what type of functions in position space that a generic rational $\mathcal{M}$ corresponds to. And in turn, although in the existing loop-level results $\mathcal{H}$ consists of generalizations of polylogarithms (the so-called multiple polylogarithms, MPL), it is also not completely known what functions their Mellin amplitudes are.

In practical computations, the above tension between the two representations further leads to two bootstrap strategies that are very different in character and that appear to gain their power in different types of questions. Regarding bootstrap strategies, in both spaces one typically starts by setting up an ansatz that shares similar pattern as in \eqref{eq:example2222p} or \eqref{eq:example2222M}, respectively. The MPLs serve as a natural basis for the position space ansatz, but it is hard to understand how they constrain the correlator from the Mellin point of view. On the other hand, by construction any Mellin space ansatz always leads to a position space function that is guaranteed to be finite at $z=\bar{z}$ (in Euclidean signature), but this condition is in some sense the hardest part to be solved in the position space bootstrap approach, due to the presence of denominators similar to that in \eqref{eq:example2222p}.  Regarding the range of applications, the Mellin space approach seems to be more powerful in the study of half-BPS operators with higher Kaluza-Klein charges \cite{Rastelli:2017udc,Alday:2020dtb,Alday:2021odx,Huang:2023ppy}, while the higher loop results are so far only obtained using the position space approach \cite{Huang:2021xws,Drummond:2022dxw,Huang:2023oxf}. Even in simple cases at tree and one loop, whether both approaches work with comparable efficiency \cite{Aprile:2017bgs,Aprile:2017qoy,Alday:2019nin}, it is fair to say that how they leads to exactly equivalent results is still not fully understood.

Due to different advantages of the two representations and their associated bootstrap approaches, one may naturally hope that a proper combination of the two will benefit us with an improved and more powerful way of bootstraping holographic correlators in general.  Yet as a preliminary step towards this goal, it is urgent that we make the equivalence between the two representations more transparent, or at least that we reduce the problem of understanding their connection to that of a much smaller class of seed functions. In fact, at tree level there exists some hints in this regard, with the help of differential operators. This idea is inspired by the recursion relation of $\bar{D}$-functions \cite{Arutyunov:2002fh}, which allows the correlator to be written as the action of a differential operator on a single function
\begin{equation}\label{eq:H2222dif}
	\mathcal{H}_{2222}(U,V) = -U^4\, \partial_U \partial_V (1+U\partial_U + V\partial_V) \bar{D}_{1111}(U,V),
\end{equation}
where
\begin{equation}\label{eq:Dbar1111UV}
	\bar{D}_{1111}(U,V) = \frac{1}{z-\zb}\left[2{\rm Li}_2(z)-2{\rm Li}_2(\zb) + \log(z\zb)\log\left(\frac{1-z}{1-\zb}\right)\right].
\end{equation}
$\bar{D}_{1111}$ plays the role of a seed function, from which the position space result \eqref{eq:example2222p} of the correlator $\mathcal{H}_{2222}$ can be derived directly by performing the differentiation. For clarity of later discussions, let us call expressions like \eqref{eq:H2222dif} a \emph{differential representation} of the correlator. The strength of this representation comes from the fact that the Mellin amplitude of $\bar{D}_{1111}(U,V)$ takes an extremely simple form, which is just a single constant! One can replace this function with its Mellin transform, similar to \eqref{eq:example2222Mt}. The differential operator now only acts on power functions of $U$ and $V$, making it straightforward to work out the corresponding Mellin amplitude \eqref{eq:example2222M}. This will be reviewed in detail in section \ref{sec:tree}.

In this paper we are going to focus on four-point half-BPS correlators in ${\rm AdS}_5 \times {\rm S}^5$ and ${\rm AdS}_5 \times {\rm S}^3$ backgrounds, which describes scattering of supergravitons and supergluons as well as their higher Kaluza-Klein modes. In these contexts we will generalize the tree-level differential representations to loop-level and stringy contributions to holographic correlators, and identify the seed functions that play analogous roles as the $\bar{D}_{1111}(U,V)$ in the above example. More explicitly, we conjecture that all these correlators can be written in a differential formalism 
\begin{equation}\label{eq:conjecture}
	\sum_i p_i(U,V,\partial_U,\partial_V) \frac{W_i(z,\zb)}{z-\zb},
\end{equation}
where the $p_i$ are polynomials on $U$, $U^{-1}$, $V$, $V^{-1}$, $\partial_U$ and $\partial_V$, and the $W_i(z,\zb)$ are certain single-valued MPLs that are parity odd under exchanging $z\leftrightarrow\zb$. In addition, the Mellin amplitudes of $\frac{W_i(z,\zb)}{z-\zb}$ are given by some simple harmonic sums (along with their generalizations). As will be illustrated later, because the differential operators $p_i(U,V,\partial_U,\partial_V)$ can always be easily translated into corresponding operations in Mellin space, the Mellin amplitudes of the correlators can be directly obtained once the Mellin counterparts of the seed functions are known. It is in this sense that the differential representation \eqref{eq:conjecture} manifestly puts the position space correlator and the Mellin amplitude on an equal footing.

It it worth to note that the existence of a differential representation of the form \eqref{eq:conjecture} is not a priori given, but we observe it holds in many examples of known holographic correlators in ${\rm AdS}_5 \times {\rm S}^5$ and ${\rm AdS}_5 \times {\rm S}^3$. In fact, the Mellin counterparts of seed functions at one loop where already noticed previously in \cite{Aprile:2020luw}, where they were used as a basis to expand the Mellin amplitudes. Yet it is still very non-trivial that such expansion can indeed be further turned into (Mellin version of) differentiations like \eqref{eq:conjecture}. Consequently, the differential representation unveils previously unknown structures of holographic correlators, and we expect that it may impose non-trivial constraints that help simplify bootstrap computations.

The rest of the paper is organized as follows. In \cref{sec:basic}, we review the basic concepts in holographic correlators that are needed for later discussions, including the definition of Mellin amplitudes, and two kinds of special functions (multiple polylogarithms and harmonic sums) appearing in the differential representations. In \cref{sec:tree}, we provide a detailed illustration on the differential representation of $\mathcal{H}_{2222}$ at the tree level. In \cref{sec:gluon2222}, we turn to the first loop-level example, the one-loop four-supergluon correlator, and use this example to show the general feature of holographic correlators in the differential representations. In \cref{sec:oneloop}, we provide more examples on one-loop correlators, including stringy corrections at the one-loop level. In \cref{sec:higherloops}, we discuss the differential representations for correlators at the two- and higher-loop level. In \cref{sec:discussion}, we provide a brief summary of our work and discuss some of the future directions.

\section{Basic concepts}\label{sec:basic}

Let us first quickly review some necessary concepts to set up the context before proceeding to our main discussions. 

\subsection{Holographic correlators and Mellin amplitudes}\label{sec:holographic}

As mentioned in the introduction we are going to focus on the super gravitons and super gluons as well as their Kaluze-Klein modes in $\mathrm{AdS}_5$ backgrounds. All these correspond to gauge-invariant half-BPS operators in the dual boundary CFTs. 

In the $\mathrm{AdS}_5\times\mathrm{S}^5$ case these operators are labeled by $\mathcal{O}_p(x,y)$ ($p=2,3,\ldots$), which has spin $\ell=0$, protected dimension $\Delta=p$, and representation $[0,p,0]$ of the $\mathrm{SU}(4)$ R-symmetry (see, e.g., the review \cite{DHoker:2002nbb}). Here $x$ denotes spacetime coordinates and $y$ is a 6-dimensional null vector encoding the R-symmetry structure. The case $p=2$ refers to the graviton and cases with $p\geq3$ its higher KK modes. As usual in CFTs, by properly splitting out an overall factor carrying conformal weights, the four-point correlator of these operators $\langle p_1p_2p_3p_4\rangle_\mathrm{GR}\equiv\langle\mathcal{O}_{p_1}(x_1,y_1)\mathcal{O}_{p_2}(x_2,y_2)\mathcal{O}_{p_3}(x_3,y_3)\mathcal{O}_{p_4}(x_4,y_4)\rangle$ relate to functions of four independent conformal-invariant cross ratios
\begin{subequations}
\begin{align}
    U&=\frac{x_{12}^2x_{34}^2}{x_{13}^2x_{24}^2}=z\bar{z},\qquad
    V=\frac{x_{14}^2x_{23}^2}{x_{13}^2x_{24}^2}=(1-z)(1-\bar{z}),\\
    \sigma&=\frac{y_{12}^2y_{34}^2}{y_{13}^2y_{24}^2}=\alpha\bar{\alpha},\qquad
    \tau=\frac{y_{14}^2y_{23}^2}{y_{13}^2y_{24}^2}=(1-\alpha)(1-\bar{\alpha}),
\end{align}
\end{subequations}
where $x_{ij}^2\equiv(x_i-x_j)^2$, $y_{ij}^2\equiv y_i\cdot y_j$. With the help of superconformal Ward identities \cite{Eden:2000bk,Nirschl:2004pa} these correlators receives the following decomposition (we abbreviate $g_{ij}\equiv y_{ij}^2/x_{ij}^2$)
\begin{equation}
    \begin{split}
        \langle p_1p_2p_3p_4\rangle_\mathrm{GR}=&g_{12}^\frac{\Sigma_{12}}{2}g_{34}^{\frac{\Sigma_{34}}{2}}\left(\frac{g_{24}}{g_{14}}\right)^{\frac{p_{21}}{2}}\left(\frac{g_{13}}{g_{14}}\right)^{\frac{p_{34}}{2}}\times\\
        &\left(\mathcal{G}_{\mathrm{GR},\{p_i\}}(z,\bar{z},\alpha,\bar{\alpha})+\frac{(z-\alpha)(z-\bar{\alpha})(\bar{z}-\alpha)(\bar{z}-\bar{\alpha})}{(z\bar{z}\alpha\bar{\alpha})^2}\mathcal{H}_{\mathrm{GR},\{p_i\}}(z,\bar{z},\alpha,\bar{\alpha})\right),
    \end{split}
\end{equation}
where $\Sigma_{ij}\equiv p_i+p_j$ and $p_{ij}\equiv p_i-p_j$. Here $\mathcal{G}$ is protected and can be fully determined in the free theory limit. So dynamics of gravitons is captured by the function $\mathcal{H}$, which is called reduced correlator and is the main object we will be studying.

In the $\mathrm{AdS}_5\times\mathrm{S}^3$ case the half-BPS operators under interest are labeled by $\mathcal{O}_p^{I}(x;v,\bar{v})$ ($p=2,3,\ldots$), which has spin $\ell=0$, dimension $\Delta=p$ \cite{Fayyazuddin:1998fb,Aharony:1998xz,Karch:2002sh,Alday:2021odx}. The gluons in $\mathrm{AdS}_5\times\mathrm{S}^3$ corresponds to $p=2$ and their higher KK modes to $p\geq3$. These operators transform in the adjoint representation of a boundary flavor group $G_F$ (indexed by $I$, which is alternatively viewed as the color for the gluons in the bulk), the spin-$\frac{p}{2}$ representation of the $\mathrm{SU}(2)_R$ R-symmetry group, and the spin-$\frac{p-2}{2}$ representation of an extra $\mathrm{SU}(2)_L$ flavor group. In our notation $v$ and $\bar{v}$ are two-component polarization vectors for $\mathrm{SU}(2)_R$ and $\mathrm{SU}(2)_L$ respectively. For the internal symmetries here we introduce two other cross ratios
\begin{equation}
    \alpha=\frac{v_{12}v_{34}}{v_{13}v_{24}},\qquad
    \beta=\frac{\bar{v}_{12}\bar{v}_{34}}{\bar{v}_{13}\bar{v}_{24}},
\end{equation}
where $v_{ij}\equiv\epsilon_{ab}v_i^av_j^b$. By similarly pulling out a proper factor and applying superconformal Ward identities \cite{Nirschl:2004pa}, we can obtain a similar decomposition of $\langle p_1p_2p_3p_4\rangle_\mathrm{YM}\equiv\langle\mathcal{O}^{I_1}_{p_1}(x_1;v_1,\bar{v}_1)\mathcal{O}^{I_2}_{p_2}(x_2;v_2,\bar{v}_2)\mathcal{O}^{I_3}_{p_3}(x_3;v_3,\bar{v}_3)\mathcal{O}^{I_4}_{p_4}(x_4;v_4,\bar{v}_4)\rangle$ 
\begin{equation}
    \begin{split}
        \langle p_1p_2p_3p_4\rangle_\mathrm{YM}=&g_{12}^\frac{\Sigma_{12}}{2}g_{34}^{\frac{\Sigma_{34}}{2}}\left(\frac{g_{24}}{g_{14}}\right)^{\frac{p_{21}}{2}}\left(\frac{g_{13}}{g_{14}}\right)^{\frac{p_{34}}{2}}\frac{1}{\bar{v}_{12}^2\bar{v}_{34}^2}\times\\
        &\left(\mathcal{G}_{\mathrm{YM},\{p_i\}}^{I_1I_2I_3I_4}(z,\bar{z},\alpha,\beta)+\frac{(z-\alpha)(\bar{z}-\alpha)}{z\bar{z}\alpha^2}\mathcal{H}_{\mathrm{YM},\{p_i\}}^{I_1I_2I_3I_4}(z,\bar{z},\alpha,\beta)\right),
    \end{split}
\end{equation}
where this time we assign $g_{ij}=v_{ij}\bar{v}_{ij}/x_{ij}^2$. Again, $\mathcal{G}$ is protected and non-trivial dynamics only enters the reduced correlator $\mathcal{H}$. There are also graviton modes in this case, but for the current exploration we only focus on the gluon sector.

We consider the weak-coupling limit of the bulk theory, and so we take the infinite limit of the size of the boundary gauge group. We also treat 't Hooft coupling to be large so as to suppress stringy effects. With this set-up we obtain a perturbative expansion of the reduced correlators for both the gravity and the gluon theories in the bulk
\begin{subequations}
\begin{align}
    \mathcal{H}_{\mathrm{GR},\{p_i\}}&=\mathcal{H}_{\mathrm{GR},\{p_i\}}^{(0)}+a_C\mathcal{H}_{\mathrm{GR},\{p_i\}}^{(1)}+a_C^2\mathcal{H}_{\mathrm{GR},\{p_i\}}^{(2)}+a_C^3\mathcal{H}_{\mathrm{GR},\{p_i\}}^{(3)}+\cdots,\\
    \mathcal{H}_{\mathrm{YM},\{p_i\}}^{\{I_i\}}&=\mathcal{H}_{\mathrm{YM},\{p_i\}}^{\{I_i\},(0)}+a_F\mathcal{H}_{\mathrm{YM},\{p_i\}}^{\{I_i\},(1)}+a_F^2\mathcal{H}_{\mathrm{YM},\{p_i\}}^{\{I_i\},(2)}+a_F^3\mathcal{H}_{\mathrm{YM},\{p_i\}}^{\{I_i\},(3)}+\cdots,
\end{align}
\end{subequations}
where the expansion parameters $a_C$ and $a_F$ are inversely proportional to the central charge and the flavor central charge in each case, respectively. These expansions have clean interpretation as diagrammatic expansions in the bulk, such that $\mathcal{H}^{(1)}$ collects contributions from tree-level Witten diagrams, $\mathcal{H}^{(2)}$ from one loop, and so on.

As mentioned in the introduction the Mellin amplitudes provide very useful representations for conformal correlators. In the current context we study Mellin amplitudes $\mathcal{M}$ for the reduced correlators $\mathcal{H}$, which are related by
\begin{equation}\label{eq:usualmellin}
\mathcal{H}_{\{p_i\}}(U,V)=\int_{-i\infty}^{+i\infty}\frac{\mathrm{d}s\,\mathrm{d}t}{(2\pi i)^2}U^{\frac{s+\mathcal{N}}{2}}V^{\frac{t-\Sigma_{23}}{2}}\mathcal{M}_{\{p_i\}}(s,t)\,\Gamma_{\{p_i\}}(s,t),
\end{equation}
where both $\mathcal{H}$ and $\mathcal{M}$ are also functions of the internal variables $\{\alpha,\bar{\alpha}\}$ or $\{\alpha,\beta\}$ (depending a the model under consideration), and the universal factor in the integrand is
\begin{equation}
\Gamma_{\{p_i\}}(s,t)=\Gamma(\frac{\Sigma_{12}-s}{2})	\Gamma(\frac{\Sigma_{34}-s}{2})\Gamma(\frac{\Sigma_{14}-t}{2})\Gamma(\frac{\Sigma_{23}-t}{2})\Gamma(\frac{\Sigma_{13}-\tilde{u}}{2})\Gamma(\frac{\Sigma_{24}-\tilde{u}}{2}),
\end{equation}
where $\tilde{u}=\Sigma-\mathcal{N}-s-t$, with $\Sigma\equiv \sum_{i=1}^4p_i$. $\mathcal{N}$ denotes the amount of supersymmetries viewed from the boundary 4d CFTs, with $\mathcal{N}=4$ in the $\mathrm{AdS}_5\times\mathrm{S}^5$ case, and $\mathcal{N}=2$ in the $\mathrm{AdS}_5\times\mathrm{S}^3$ case.

For the convenience and clarity of discussions in this paper, it is preperrable to switch to a new set of Mellin-Mandelstam variables instead of the usual $\{s,t,\tilde{u}\}$. Let us denote
\begin{equation}
    s_0=\max(\Sigma_{12},\Sigma_{34}),\quad t_0=\max(\Sigma_{14},\Sigma_{23}),\quad
    \tilde{u}_0=\max(\Sigma_{13},\Sigma_{24}).
\end{equation}
Then the new variables are defined by
\begin{equation}
S=\frac{s-s_0}{2},\qquad
T=\frac{t-t_0}{2},\qquad
\tilde{U}=\frac{\tilde{u}-\tilde{u}_0  }{2},
\end{equation}
such that they satisfy the relation
\begin{equation}
S+T+\tilde{U}=\frac{1}{2}(\Sigma-s_0-t_0-\tilde{u}_0-\mathcal{N})\equiv-\mathcal{E}-\frac{\mathcal{N}}{2}.
\end{equation}
Here $\mathcal{E}$ is called extremality, which is a useful characterization of structural complexity of the correlator, and it takes integral values and starts at $\mathcal{E}=2$ for the simplest non-trivial correlators. With these new variables the $\Gamma$ factors become
\begin{equation}\label{eq:Gammacanonical}
\Gamma_{\{p_i\}}(s,t)\equiv\widetilde{\Gamma}_{\{p_i\}}(S,T)=\Gamma(-S)\Gamma(\tmp_s-S)\Gamma(-T)\Gamma(\tmp_t-T)\Gamma(-\tilde{U})\Gamma(\tmp_u-\tilde{U}),
\end{equation}
where
\begin{equation}
    \tmp_s=\frac{|\Sigma_{12}-\Sigma_{34}|}{2},\quad  \tmp_t=\frac{|\Sigma_{14}-\Sigma_{23}|}{2},\quad  \tmp_u=\frac{|\Sigma_{13}-\Sigma_{24}|}{2}.
\end{equation}
Without loss of generality we can always assume $\Sigma_{23}\geq\Sigma_{14}$ (and the other correlators are related by crossing), then we can rewrite the relation \eqref{eq:usualmellin} into
\begin{equation}\label{eq:MtoHnew}
\mathcal{H}_{\{p_i\}}(U,V)=\int_{-i\infty}^{+i\infty}\frac{\mathrm{d}S\mathrm{d}T}{(2\pi i)^2}U^{S+\frac{s_0+\mathcal{N}}{2}}V^{T}\widetilde{\mathcal{M}}_{\{p_i\}}(S,T)\,\widetilde{\Gamma}_{\{p_i\}}(S,T),
\end{equation}
where $\widetilde{\mathcal{M}}(S,T)=4\mathcal{M}(s,t)$. In this new convention, e.g., the 4-graviton Mellin amplitude \eqref{eq:example2222M} turns into
\begin{equation}\label{eq:M2222ST}
\widetilde{\mathcal{M}}_{2222}(S,T)=\frac{1}{(1+S)(1+T)(1+\tilde{U})}\equiv\frac{1}{(1+S)(1+T)(-3-S-T)}\,.
\end{equation}

\subsection{Functions in position space and Mellin space}

In the introduction we briefly mentioned that in general the function space of the holographic correlators are not fully understood in either position space or Mellin space. Nevertheless, existing literature up to one loop suggest a proper candidate on each side to start with.

\subsubsection{Multiple polylogarithms}

In the position space the known results up to one loop for four-point correlators in both $\mathrm{AdS}_5\times\mathrm{S}^5$ and $\mathrm{AdS}_5\times\mathrm{S}^3$ cases manifest as linear combinations of the so-called multiple polylogarithms (MPLs), which are generalizations of the logarithm function. Each of such functions comes with a vector of parameters $\vec{a}\equiv\{a_1,a_2,\ldots,a_n\}$ and is denoted as $G_{\vec{a}}(z)$ (or simply $G_{\vec{a}, z}$). They can be recursively defined in terms of iterated integrals \cite{Goncharov:1998kja,Goncharov:2001iea}
\begin{equation}
	G_{a_1a_2\ldots a_n}(z)=\int_0^z\frac{\mathrm{d}t}{t-a_1}G_{a_2a_3\ldots a_n}(t)\,\qquad G(z)=1\,,
\end{equation}
with the special case for $n$ zero parameters $\vec{0}_n\equiv\{0,0,\ldots,0\}$
\begin{equation}
	G_{\vec{0}_n}(z)=\frac{1}{n!}\log^nz\,.
\end{equation}
Size of the vector $\vec{a}$ is called the (transcendental) weight of the MPL. Simple examples are the classical polylogarithms $\mathrm{Li}_n(z)=-G_{0\ldots01}(z)$. In terms of these functions $\bar{D}_{1111}(U,V)$ in \eqref{eq:Dbar1111UV} can be written as 
\begin{equation}
	\bar{D}_{1111}(U,V)=\frac{G_0(\bar{z})G_1(z)-G_0(z)G_1(\bar{z})-G_{01}(z)+G_{01}(\bar{z})+G_{10}(z)-G_{10}(\bar{z})}{z-\bar{z}}\,.
\end{equation}

In fact, the four-point functions only have singularities at $z=0,1,\infty$ (and similar for $\bar{z}$), which forces elements in the weight vector $\vec{a}$ to be either $0$ or $1$ in most situations. This belongs to a special subclass of MPLs called harmonic polylogarithms (HPLs) \footnote{The harmonic polylogarithms also allow elements in $\vec{a}$ to be valued at $-1$, but this case is not observed to be relevant in this study.}. At weight three or higher one may also need extra singluarities in $G$ at $z=\bar{z}$ (though this is not seen on the principal sheet) \cite{Drummond:2019hel,Huang:2021xws}, such that $z$ or $\bar{z}$ are also need in the weight vector, which we will see in explicit examples later. 

In addition, the correlators have to be single-valued in the Euclidean region, where $\bar{z}=z^*$. Hence very naturally they are built from single-valued combinations of MPLs (SVMPLs) \cite{Brown:2004ugm}. Given the above constraints on the potential singularities, there is a well-defined algorithm to construct the set of independent SVMPLs at each weight \cite{Chavez:2012kn}, upon which the correlator can be decomposed. Alternatively, given each MPL $G_{\vec{a}}(z)$ there always corresponds an SVMPL $G_{\vec{a}}^{\rm sv}(z,\bar{z})$ via a method called single-valued projection \cite{Brown:2013gia,Brown:2015ylf}. As we will see, this allows to express many results in compact forms. Take $\bar{D}_{1111}(U,V)$ again as an example, we have
\begin{equation}
	\bar{D}_{1111}(U,V)=\frac{G_{10}^{\rm sv}(z,\bar{z})-G_{01}^{\rm sv}(z,\bar{z})}{z-\bar{z}}\,,
\end{equation}
where the single-valued projection dictates that (one can conveniently generate these using \texttt{PolyLogTools} \cite{Duhr:2019tlz})
\begin{subequations}
\begin{align}
	G_{10}^{\rm sv}(z,\bar{z})&=G_0(\bar{z})G_1(z)+G_{10}(z)+G_{01}(\bar{z})\,,\\
	G_{01}^{\rm sv}(z,\bar{z})&=G_0(z)G_1(\bar{z})+G_{10}(\bar{z})+G_{01}(z)\,.
\end{align}
\end{subequations}

We can further decompose SVMPLs into two categories according to their parity under exchanging $z\leftrightarrow\zb$. For an arbitrary SVMPL $G(z,\zb)$, it can always be decomposed into $G(z,\bar{z})+G(\zb,z)$ and $G(z,\bar{z})-G(\zb,z)$, with the former parity even and the latter parity odd. The parity odd sector will play an important role in the differential representation. For instance, one can immediately check the combination $G_{10}^{\rm sv}(z,\bar{z})-G_{01}^{\rm sv}(z,\bar{z})$ appearing in $\bar{D}_{1111}$ is in the parity odd sector, since there will be a minus sign after taking $z\leftrightarrow\zb$.

\subsubsection{Harmonic sums}
It is known that the ordinary single-variable Mellin transform of HPLs belong to another class of objects called harmonic sums \cite{Vermaseren:1998uu}. Although the Mellin transform relations \eqref{eq:MtoHnew} used in the four-point correlators is not exactly the same as transforming individual HPLs, one may not be surprised that these objects also make an appearance in the Mellin amplitudes. As will be presented later, this is indeed the case at least at one loop. The harmonic sums are originally defined in terms of iterated sums of the form
\begin{equation}
	S_{a_1,a_2,\ldots,a_n}(n)=\sum_{i=1}^n\frac{1}{i^{a_1}}S_{a_2,a_3,\ldots,a_n}(i)\,,
\end{equation}
where $a_i$ are positive integers.\footnote{More generally the harmonic sums can also be defined with negative integral $a_i$, but these are not relevant here.} The weight of harmonic sums, similarly, is defined to be the sum of $a_i$.

The simplest cases of them, $S_a(n)$, are the harmonic numbers, and they are related to the polygamma functions
\begin{equation}
	\psi^{(\nu)}(x)=\frac{\mathrm{d}^\nu\psi(x)}{\mathrm{d}x^\nu}\,,\qquad\psi^{(0)}(x)\equiv\psi(x)=\frac{\mathrm{d}\log \Gamma(x)}{\mathrm{d}x}
\end{equation}
by the identity
\begin{equation}
	S_a(n)=\frac{(-1)^{a-1}}{(a-1)!}\left(\psi^{(a-1)}(n+1)-\psi^{(a-1)}(1)\right)\,.
\end{equation}
Note that the polygamma $\psi^\nu(x)$ is defined on the whole complex plane of $x$ except at $x=-1,-2,-3,\ldots$ where it has poles. This gives rise to the analytic continuation of harmonic numbers beyond positive integral $n$. In particular, their Laurent expansion at the poles are in general
\begin{equation}
	S_a(x)=\frac{-1}{(x+p)^a}+\mathcal{O}(1)\,,\qquad p=-1,-2,-3,\ldots\,.
\end{equation}
This is important for our purpose, since very often our first understanding of Mellin amplitudes comes with the information about their poles and residues. With the help of the Laurent expansion like above, we will be able to learn what the exact function potentially is. Analytic continuations of arbitrary harmonic sums can be systematically worked out with the help of oridinary Mellin transform \cite{Velizhanin:2020avm}. In general they also have poles at non-positive integers. Here we just list the Laurent expansion of several simplest cases up to the linear order
\begin{subequations}
\begin{align}
	\label{eq:LaurentS1}S_1(x)&=\frac{-1}{x+p}+S_1(p-1)+(S_2(p-1)+\zeta_2)(x+p)+\mathcal{O}((x+p)^2)\,,\\
	S_2(x)&=\frac{-1}{(x+p)^2}-S_2(p-1)+2(-S_3(p-1)+\zeta_3)(x+p)+\mathcal{O}((x+p)^2)\,,\\
	S_{1,1}(x)&=\frac{-S_1(p-1)}{x+p}+\Big(-2S_2(p-1)+S_{1,1}(p-1)-\zeta_2\Big)+\Big(-3S_3(p-1)\nonumber\\
	&\quad+S_{1,2}(p-1)+S_{2,1}(p-1)+S_1(p-1)\zeta_2+2\zeta_3\Big)(x+p)+\mathcal{O}((x+p)^2)\,.
\end{align}
\end{subequations}

\section{Tree level: an illustration}\label{sec:tree}

Despite of the explicit integral relation between Mellin ampliutdes and the reduced correlator \eqref{eq:MtoHnew}, it is not always necessary to work through this relation in order to verify the equivalence between the two types of observables. To warm up, let us first have a look at the tree-level case, where the problem boils down to that of a single seed function $\bar{D}_{1111}(U,V)$, whose position space expression is already presented in \eqref{eq:Dbar1111UV}. This function arises from a scalar contact diagram and its Mellin representation reads
\begin{equation}\label{eq:Dbar1111original}
	\bar{D}_{1111}(U,V)=\int\frac{\mathrm{d}S\mathrm{d}T}{(2\pi i)^2}U^SV^T\Gamma(-S)^2\Gamma(-T)^2\Gamma(1+S+T)^2,
\end{equation}
such that its Mellin amplitude is trivially a constant $1$. The Mellin transformation here,
\begin{equation}\label{eq:MellinStandardForm}
    \int\frac{\mathrm{d}S\mathrm{d}T}{(2\pi i)^2}U^S V^T (\cdots) \Gamma(-S)^2\Gamma(-T)^2\Gamma(1+S+T)^2,
\end{equation}
are called the {\it standard form} of Mellin transformation in this paper. We also denote the corresponding $\Gamma$ factors to be unlabeled
\begin{equation}\label{eq:Gammastandard}
    \widetilde{\Gamma}(S,T)\equiv \Gamma(-S)^2\Gamma(-T)^2\Gamma(1+S+T)^2.
\end{equation}
In this simple example the above integral relation \eqref{eq:Dbar1111original} can be easily verified. On the one hand, one can Taylor-expand the LHS around $U=0$ and $V=0$, yielding
\begin{equation}
	\begin{split}
		\bar{D}_{1111}=&((1+V+V^2+\cdots)+(1+4V+9V^2+\cdots)U+\cdots)\log U\log V\\
		&+(\cdots)\log U+(\cdots)\log V+(\cdots).
	\end{split}
\end{equation}
On the other hand, on the RHS one can also bend the $S$ and $T$ contour to the right, so as to turn the integral into summation over residues of the integrand at poles $S=m$ ($m\in\mathbb{N}$) and $T=n$ ($n\in\mathbb{N}$), and this produces exactly the same expansion. 

It is generally expected that arbitrary tree-level correlator in the theories we consider can be obtained by acting certain differential operator on $\bar{D}_{1111}$, based on the facts that they all receive a decomposition onto finite number of contact diagrams and that any contact diagram is recursively related to $\bar{D}_{1111}$ by differentiation. The virtue of such differential representation is that they can be mapped to some corresponding actions in Mellin space. Starting with any correlator $\mathcal{H}_{\{p_i\}}(U,V)$ with its Mellin amplitude $\widetilde{\mathcal{M}}_{\{p_i\}}(S,T)$, let us consider the following operations (for convenience we omit the labels $\{p_i\}$ in $\mathcal{H}$ and $\widetilde{\mathcal{M}}$ below). 
\begin{itemize}
\item We can always rewrite the Mellin transformation \eqref{eq:MtoHnew} of $\mathcal{H}(U,V)$ into the standard form by the recursion relations of $\Gamma$ functions 
\begin{equation}\label{eq:Mtilde2Standard}
	\begin{split}
        \mathcal{H}(U,V)&=\int\frac{\mathrm{d}S\mathrm{d}T}{(2\pi i)^2}U^{S+\frac{s_0+\mathcal{N}}{2}}V^{T}\widetilde{\mathcal{M}}(S,T)\,\widetilde{\Gamma}_{\{p_i\}}(S,T)\\
        &= U^{\frac{s_0+\mathcal{N}}{2}} \int\frac{\mathrm{d}S\mathrm{d}T}{(2\pi i)^2}U^{S}V^{T}\,  \widetilde{\Gamma}(S,T)\, \widetilde{\mathcal{M}}(S,T)\, (-S)_{\tmp_s}(-T)_{\tmp_t}\\
        &\qquad \times   (1+S+T)_{\mathcal{E}+\frac{\mathcal{N}}{2}-1}(1+S+T)_{\mathcal{E}+\frac{\mathcal{N}}{2}+\tmp_u-1},
    \end{split}
\end{equation}
where $(a)_n\equiv\frac{\Gamma(a+n)}{\Gamma(a)}$ is the Pochhammer symbol.  In this way, the Mellin amplitude of $U^{-\frac{s_0+\mathcal{N}}{2}}\mathcal{H}(U,V)$ in the standard form becomes
\begin{equation}\label{eq:standardMellin}
    \widetilde{\mathcal{M}}(S,T)\, (-S)_{\tmp_s}(-T)_{\tmp_t} (1+S+T)_{\mathcal{E}+\frac{\mathcal{N}}{2}-1}(1+S+T)_{\mathcal{E}+\frac{\mathcal{N}}{2}+\tmp_u-1}.
\end{equation}
We call \eqref{eq:standardMellin} the {\it standard Mellin amplitude} of the correlator. In general, $\tmp_{s,t,u}$ and $\mathcal{E}+\frac{\mathcal{N}}{2}-1$ are always positive, so there will not be additional poles in the standard Mellin amplitude comparing with the original one. Instead, some poles in $\widetilde{\mathcal{M}}(S,T)$ may be canceled by the extra Pochhammer symbols.

\item When multiplying some integral powers $U^a$ to the function in the standard form, in order to keep the transformation standard, we have
\begin{equation}
	\begin{split}
		U^a\mathcal{H}(U,V)&=\int\frac{\mathrm{d}S\mathrm{d}T}{(2\pi i)^2}U^{S}V^{T}\widetilde{\mathcal{M}}(S-a,T)\,\widetilde{\Gamma}(S-a,T)\\
		&=\int\frac{\mathrm{d}S\mathrm{d}T}{(2\pi i)^2}U^{S}V^{T}\,\widetilde{\Gamma}(S,T)\, \big[(-S)_{a}(1+S+T)_{-a}\big]^2\, \widetilde{\mathcal{M}}(S-a,T)\, ,
	\end{split}
\end{equation}
where in the first line we redefined the variable $S\mapsto S-a$ in order to preserve the $U$ factor, and in the second line we applied recursion relations for $\Gamma$ functions to bring the $\Gamma$ factors back to the standard one \eqref{eq:Gammastandard}. Therefore the correspondence $\mathcal{H}(U,V)\Leftrightarrow\widetilde{\mathcal{M}}(S,T)$ becomes
\begin{equation}\label{rule:U}
	U^a\mathcal{H}(U,V)\quad\Leftrightarrow\quad \big[(-S)_{a}(1+S+T)_{-a}\big]^2\, \widetilde{\mathcal{M}}(S-a,T)
\end{equation}
for the multiplication action. By similar reasoning we also have
\begin{equation}\label{rule:V}
	V^b\mathcal{H}(U,V)\quad\Leftrightarrow\quad \big[(-T)_{b}(1+S+T)_{-b}\big]^2\, \widetilde{\mathcal{M}}(S,T-b)\, .
\end{equation}
It is worth noting that depending on the sign of $a$ and $b$, different denominators will be generated in this process. For example, when $a>0$, the factor $\big[(-S)_{a}(1+S+T)_{-a}\big]^2$ becomes
\begin{equation}
    \frac{(-S)^2(-S+1)^2\cdots(-S+a-1)^2}{(S+T)^2(-1+S+T)^2\cdots(1-a+S+T)^2},
\end{equation}
while for $a<0$, it is 
\begin{equation}
    \frac{(1+S+T)^2(2+S+T)^2\cdots(-a+S+T)^2}{(-S-1)^2(-S-2)^2\cdots(-S+a)^2}.
\end{equation}
\item When taking derivative $\partial_U$, without loss of generality we can alternatively consider acting with $ \mathcal{D}_U\equiv U\partial _U$ (since $\partial_U=(U^{-1})(U\partial_U)$ and the action of $U^{-1}$ has been analyzed above). This operation has a simple consequence on the Mellin ampliutdes
\begin{equation}
	\begin{split}\label{rule:UdU}
		\DU \mathcal{H}(U,V)=\int\frac{\mathrm{d}S\mathrm{d}T}{(2\pi i)^2}U^{S}V^{T}\,\widetilde{\Gamma}(S,T)\times S\,\widetilde{\mathcal{M}}(S,T)\,.
	\end{split}
\end{equation}
By applying this action multiple times we then have another correspondence
\begin{equation}\label{rule:VdV}
	(\DU)^a\,\mathcal{H}(U,V)
	\quad\Leftrightarrow\quad S^a\widetilde{\mathcal{M}}(S,T)\,,
\end{equation}
and similarly
\begin{equation}\label{rule:VVdVV}
	(\DV)^b\,\mathcal{H}(U,V)
	\quad\Leftrightarrow\quad T^b\widetilde{\mathcal{M}}(S,T)\,.
\end{equation}
\end{itemize}
From the above discussions it is obvious that, as long as we start with a rational function for the Mellin amplitude, then any correlator derived by composing the above operations will again has a rational Mellin amplitude. This is what happens in all tree-level scattering in the theories under consideration.

As an explicit example let us check the simplest case $\mathcal{H}_{2222}$ of gravitons, whose Mellin amplitude was shown in \eqref{eq:M2222ST}. Recall our conventions on the standard Mellin amplitude \eqref{eq:standardMellin}, we had better consider $U^{-\frac{s_0+\mathcal{N}}{2}}\mathcal{H}_{2222} \equiv U^{-4} \mathcal{H}_{2222}$ at the beginning. Given the differential relation \eqref{eq:H2222dif} between $\mathcal{H}_{2222}$ and $\bar{D}_{1111}$, the differential operator there can be organized as a sequence of five operations
\begin{equation}
	 U^{-4} \mathcal{H}_{2222}=(U^{-1})(V^{-1})\underbrace{(-\DU)(\DV)(1+\DU+\DV)}_{\partial^3}\bar{D}_{1111}\,.
\end{equation}
The first three operations act purely as multiplications in Mellin space. So starting from the Mellin counterpart for $\bar{D}_{1111}$ the rules \eqref{rule:U} to \eqref{rule:VdV} gives
\begin{equation}
	\begin{split}
		1\,\underset{\partial^3}{\longrightarrow}\,&-S\,T\,(1+S+T)\\
		\underset{V^{-1}}{\longrightarrow}\,&-\left(\frac{1+S+T}{-T-1}\right)^2 S\,(T+1)\,(2+S+T)\\
		\underset{U^{-1}}{\longrightarrow} \,&\, -\left(\frac{1+S+T}{-S-1}\right)^2\left(\frac{2+S+T}{-T-1}\right)^2 (S+1)\,(T+1)\,(3+S+T)\\
		& = -\frac{1}{(1+S)(1+T)(3+S+T)} (1+S+T)^2(2+S+T)^2(3+S+T)^2 \,.
	\end{split}
\end{equation}
Hence we directly obtain the Mellin representation of $U^{-4} \mathcal{H}_{2222}$ in the standard form
\begin{equation}
    U^{-4} \mathcal{H}_{2222} = \int\frac{\mathrm{d}S\mathrm{d}T}{(2\pi i)^2} U^S V^T \left[-\frac{(1+S+T)^2(2+S+T)^2(3+S+T)^2}{(1+S)(1+T)(3+S+T)}\right]  \widetilde{\Gamma}(S,T)\, .
\end{equation}
It is straightforward to further work out the Mellin amplitude $\widetilde{\mathcal{M}}_{2222}(S,T)$ presented in \eqref{eq:M2222ST}, by recognizing $(1+S+T)^2(2+S+T)^2(3+S+T)^2\,\widetilde{\Gamma}(S,T)\equiv\widetilde{\Gamma}_{2222}(S,T)$ to be the traditional definition \eqref{eq:Gammacanonical} of $\Gamma$ factors in the Mellin amplitudes.

We can also reverse the analysis and straightforwardly reconstruct the differential operator when the target Mellin amplitude is given. The crucial point here is that only the multiplication rules \eqref{rule:U} and \eqref{rule:V} have a chance to create factors in the denominator. This can be used  as a guideline to read off the operations in need. Again taking the above example as an illustration, we seek for a proper sequence of operations that transform
\begin{equation}
	1\,\longrightarrow\,-\frac{(1+S+T)^2(2+S+T)^2(3+S+T)}{(1+S)(1+T)}
\end{equation}
in Mellin space. We can start by thinking about multiplying by some factor $U^aV^b$. Since we need to create a $(1+S)$ and $(1+T)$ factor in the denominator, a minimal assumption is to use $U^{-1}V^{-1}$. This yields
\begin{equation}
	1\,\underset{U^{-1}V^{-1}}{\longrightarrow}\,\frac{(1+S+T)^2(2+S+T)^2}{(-S-1)^2(-T-1)^2}\,.
\end{equation}
The remaining task is to fix the wrong powers of denominator factors and the numerators by $\DU$ and $\DV$, which can be done by applying the differential operator $-(1+\DU)(1+\DV)(3+\DU+\DV)$. Consequently we find that
\begin{equation}
	\mathcal{H}_{2222}(U,V)=-U^4(1+\DU)(1+\DV)(3+\DU+\DV)\left(U^{-1}V^{-1}\bar{D}_{1111}\right).
\end{equation}
One can explicictly verify that the differential operator here exactly matches that one in \eqref{eq:H2222dif}. Of course they are expressing the same differential operator but in different forms. While there are in general many ways to write out a differential operator, it is guaranteed that both they and their corresponding Mellin space operations lead to the same results.

By now we can appreciate that, once we confirm that a class of functions arise from some seed functions by differentiation, then the connections between them and their Mellin amplitudes can all be reduced to that of the seed function and its Mellin counterpart. In the case of tree-level amplitudes of super gravitons it is well-known that any $\mathcal{H}_{p_1p_2p_3p_4}$ can be related to $\bar{D}_{1111}$ by certain differential operators, and the explicity results can be more systematically worked out with the help of the conjectured hidden 10d conformal symmetries observed recently \cite{Caron-Huot:2018kta}. This works exactly in the same way for the 4-point correlators of half-BPS operators in the $\mathrm{AdS}_5\times\mathrm{S}^3$ case as well, given the tree-level results and the associated 8d hidden conformal symmetries in \cite{Alday:2021odx}. Hence for this class of tree-level correlators there is indeed a unique seed function $\bar{D}_{1111}$. It is helpful to learn about the seed functions appearing at loop level as well, in order to further simplify the connection between position space correlators and the Mellin amplitudes.

\section{$\langle 2222 \rangle_\mathrm{YM}$ in ${\rm AdS}_5\times {\rm S}^3$}\label{sec:gluon2222}

Now we turn to the simplest one-loop correlator in ${\rm AdS}_5\times {\rm S}^3$, which is the correlator of four supergluons $\langle 2222 \rangle_{\rm YM}$. In this case, we will begin our analysis of the correlator from its Mellin amplitude and try to find the differential representation of this correlator. 

Due to the presence of the color group $G_F$ in the bulk theory, the discussion on the one-loop Mellin amplitude in this case can be simplified by first carrying out a decomposition onto three independent color factors
\begin{equation}
	\widetilde{\mathcal{M}}^{\{I_i\},(2)}_{\mathrm{YM},2222} = \dst \widetilde{\mathcal{M}}^{(2),st}_{\mathrm{YM},2222}(S,T) + \dsu \widetilde{\mathcal{M}}^{(2),su}_{\mathrm{YM},2222}(S,\tilde{U}) + \dtu \widetilde{\mathcal{M}}^{(2),tu}_{\mathrm{YM},2222}(T,\tilde{U}),
\end{equation}
where (using structure constants $f^{IJK}$)
\begin{subequations}
\begin{align}
    \mathtt{d}_{st}&=f^{JI_1K}f^{KI_2L}f^{LI_3M}f^{MI_4J},\\
    \mathtt{d}_{su}&=f^{JI_1K}f^{KI_2L}f^{LI_4M}f^{MI_3J},\\
    \mathtt{d}_{tu}&=f^{JI_1K}f^{KI_3L}f^{LI_2M}f^{MI_4J}.
\end{align}
\end{subequations}
Due to the crossing symmetry of the full correlator, the three partial amplitudes $\widetilde{\mathcal{M}}^{(2),st}_{\mathrm{YM},2222}$, $\widetilde{\mathcal{M}}^{(2),su}_{\mathrm{YM},2222}$ and $\widetilde{\mathcal{M}}^{(2),tu}_{\mathrm{YM},2222}$ are in fact the same function, but with different variables. Hence it suffices to just consider the one in $\dst$ channel here. Since in each section we will only consider one correlator, for simplicity we will always drop the YM/GR and $\{p_i\}$ labels on the Mellin amplitudes in this section and the following section. The Mellin amplitude $\widetilde{\mathcal{M}}^{(2)}(S,T)$ is given by
\begin{equation}
	\widetilde{\mathcal{M}}^{(2)}(S,T) = \sum_{m,n=0}^{\infty} \frac{a_{m,n}}{(S-m)(T-n)},\label{eq:gluon2222oneloop}
\end{equation}
with the coefficients \cite{Alday:2021ajh}
\begin{equation}\label{eq:YM2222stamn}
	a_{m,n} = \frac{3 m^2 n+2 m^2+3 m n^2+8 m n+3 m+2 n^2+3 n}{3(m+n) (m+n+1) (m+n+2)}.
\end{equation}
It is related to the position space correlator in the $\dst$ channel by
\begin{equation}\label{eq:MT2222YM}
	\mathcal{H}^{(2)}_{st,2222}(U,V) = \int \frac{\dd S \dd T}{(2\pi i)^2}U^{S+3}V^{T} \widetilde{\mathcal{M}}^{(2)}(S,T) \Gamma^2(-S)\Gamma^2(-T)\Gamma^2(3+S+T).
\end{equation}

\subsection{The resummation of Mellin amplitudes}\label{sec:resum}

In preparation for working out the differential representation, it is helpful to first have a closed-form expression for the Mellin amplitude $\widetilde{\mathcal{M}}^{(2)}$. The formal summation \eqref{eq:gluon2222oneloop} turns out to be divergent, but there are various methods to regularize it. We firstly adopt the method described in section 5.2 in \cite{Alday:2021ajh}, which involves taking derivatives with respect to $S$ and $T$ to render the sum convergent. 

As a warm-up, let us consider a simpler divergent sum with only one single variable
\begin{equation}\label{eq:sumex1}
	f_1(S) = \sum_{m=0}^\infty\frac{1}{S-m}.
\end{equation}
This sum diverges in a similar way to the harmonic series and is not well-defined. However, by taking derivatives with respect to $S$, the sum becomes convergent and can be expressed in closed form 
\begin{equation}
	\partial_S f_1(S) = -\sum_{m=0}^\infty \frac{1}{(S-m)^2} = -\psi'(-S).
\end{equation}
Therefore, we can integrate over $S$ and define the original sum to be 
\begin{equation}\label{eq:f1ClosedForm}
	f_1(S) = \psi(-S) + C_0 = S_1(-S-1) + C
\end{equation}
up to an ambiguous constant $C$. Similarly, when the residue at $S=m$ is a polynomial $p(m)$ with degree $k$, the divergent sum
\begin{equation}\label{eq:sumex2}
	f_2(S) = \sum_{m=0}^\infty \frac{p(m)}{S-m}
\end{equation}
can be evaluated by taking $(k+1)$-th derivatives with respect to $S$ and integrating it back 
\begin{equation}
	f_2(S) = p(S)S_1(-S-1) + q(S),
\end{equation}
where $q(S)$ is an ambiguous degree-$k$ polynomial. Similar ambiguous polynomial will also appear in the divergent sum of Mellin amplitudes. They have specific physical meanings, corresponding to the ambiguities arising from the renormalization of UV divergences at the loop level \cite{Alday:2019nin}. For supergluons, the ambiguities can only be a constant.

There is another way to think about these divergent sums, which will provide a more convenient way for our computation. The idea is that, for divergent sums like \eqref{eq:sumex1} and \eqref{eq:sumex2}, we should not literally sum them up, but should view them as formal expressions indicating the poles and residues of certain functions. When the behavior of these functions at infinity are specified, we can uniquely reconstruct the functions from poles using dispersion relations. However, the behavior at infinity is usually not indicated in the sum, so we have to leave it ambiguous. This explains the polynomial ambiguities mentioned earlier, which correspond to all possible subtractions in the dispersion relations. This viewpoint works well for divergent sums like \eqref{eq:sumex1} and \eqref{eq:sumex2}, and it is also suitable for Mellin amplitudes since the physical information of Mellin amplitudes is fully encoded in their poles, residues, and behavior at infinity (also known as the Regge behavior) \cite{Penedones:2019tng}.

For convergent sums, the above perspective often allows us to determine the final result based on some guesswork, without explicitly summing infinitely many terms. One simple example is to consider
\begin{equation}\label{eq:sumex3}
	f_3(S) = \sum_{m=0}^{\infty}\frac{1}{m+a}\frac{1}{S-m}.
\end{equation}
We are going to find a function with the same poles and residues as the sum. Note that without the factor $(m+a)^{-1}$ the above poles have the same structure as those of the harmonic sum $S_1(-S-1)$, c.f.~\eqref{eq:LaurentS1}. So a tempting guess is obtained by directly replacing $m$ in the residue $(m+a)^{-1}$ back to $S$
\begin{equation}
	\frac{S_1(-S-1)}{S+a}.
\end{equation}
This however introduces an additional pole at $S=-a$. To eliminate this unwanted pole, we need to manually cancel its residue, which can be done by
\begin{equation}
	f_3(S) = \frac{S_1(-S-1)-S_1(a-1)}{S+a}.
\end{equation}
This indeed yields the same result as that from summing \eqref{eq:sumex3}. 

A more non-trivial example which is more relevant to the one-loop Mellin amplitude $\widetilde{\mathcal{M}}^{(2)}$ is the double sum
\begin{equation}
	f_4(S,T)=\sum_{m,n=0}^{\infty}\frac{1}{m+n+1}\frac{1}{(S-m)(T-n)}.
\end{equation}
Once again, by comparing the structure of $S$ poles and $T$ poles respectively with available harmonics sums, we draw the following guess
\begin{equation}
	\frac{S_1(-S-1)S_1(-T-1)}{S+T+1},
\end{equation}
which introduces a new $S$ pole at $S=-T-1$ with residue $S_1(T)S_1(-T-1)$. To cancel this additional pole, we can make used of an identity of harmonic sums \footnote{A proof of this identity is provided in \cref{appx:identity}.}
\begin{equation}\label{eq:HSidentity}
	S_1(T)S_1(-T-1)-S_{1,1}(T)-S_{1,1}(-T-1)-2\zeta_2 = 0.
\end{equation}
This suggests the final result should be 
\begin{equation}\label{eq:sumres2}
	f_4(S,T)=\frac{S_1(-S-1)S_1(-T-1)-S_{1,1}(-S-1)-S_{1,1}(-T-1)-2\zeta_2}{S+T+1}.
\end{equation}
Again, this expression yields the correct result, as can be verified using Mathematica. As we will see, the particular combination \footnote{This function is equivalent to the reduced kernel $\mathcal{K}[z,\zb]$ found in \cite{Aprile:2020luw}. }
\begin{equation}
	\Phi(S,T) = S_1(-S-1)S_1(-T-1)-S_{1,1}(-S-1)-S_{1,1}(-T-1)-2\zeta_2
\end{equation}
turns out to be the Mellin amplitude of a seed function for one-loop correlators in the differential representation.

Using \eqref{eq:sumres2} as a building block, we can easily construct the closed form expression of $\widetilde{\mathcal{M}}^{(2)}(S,T)$ as well as other one-loop Mellin amplitudes. For $\widetilde{\mathcal{M}}^{(2)}(S,T)$, by rewriting the residues $a_{m,n}$ in \eqref{eq:YM2222stamn} using partial fractions with respect to $m$,
\begin{equation}
	a_{m,n} = -\frac{2 n^2}{3 (m+n)}+\frac{n^2+n+1}{3 (m+n+1)}+\frac{(n+1)^2}{3 (m+n+2)},
\end{equation}
we decompose the sum into three parts. Applying \eqref{eq:sumres2}, the middle part yields
\begin{equation}
\begin{split}
    &\sum_{m,n=0}^\infty\frac{n^2+n+1}{3}\left(\frac{1}{m+n+1}\frac{1}{(S-m)(T-n)}\right)\\
	&\longrightarrow\frac{T^2+T+1}{3}f_4(S,T)\equiv\frac{T^2+T+1}{3(S+T+1)}\Phi(S,T),
\end{split}
\end{equation}
while the first and last parts can be computed similarly by shifting the definition of $T$ and $n$. These three parts give
\begin{equation}\label{eq:2222YMtmp}
	-\frac{2 T^2 }{3 (S+T)}\Phi(S,T-1) + \frac{\left(T^2+T+1\right) }{3 (S+T+1)}\Phi(S,T)+\frac{(T+1)^2 }{3 (S+T+2)}\Phi(S,T+1).
\end{equation}
However, we need to be cautious with this result, because the presence of extra numerators in $n$ indicates that each part is obtained from a divergent sum over $n$. This may potentially introduce ambiguous terms regular in $T$. Notice that the original sum \eqref{eq:gluon2222oneloop} is actually convergent in $n$ summation for each fixed $S$ pole (i.e., for fixed $m$), so there will not be any ambiguities when considering the function on $S$ poles. Comparing \eqref{eq:2222YMtmp} for poles at $S=m$ with the partial sum over $n$ in \eqref{eq:gluon2222oneloop}, we find that the difference is $-1$. Following \eqref{eq:f1ClosedForm}, this can be compensated by $-S_1(-S-1)-C$, leading to the final result
\begin{align}\label{eq:gluonresum2222}
	\widetilde{\mathcal{M}}^{(2)}(S,T) =& -\frac{2 T^2 }{3 (S+T)}\Phi(S,T-1) + \frac{\left(T^2+T+1\right) }{3 (S+T+1)}\Phi(S,T) \nonumber \\
	& +\frac{(T+1)^2 }{3 (S+T+2)}\Phi(S,T+1)- S_1(-S-1)-C.
\end{align}
Here $C$ is an arbitrary constant that accounts for regularization of the UV divergence in $\widetilde{\mathcal{M}}^{(2)}(S,T)$. Note that this result may not appear symmetric in $S$ and $T$, but upon closer inspection, we can verify that \eqref{eq:gluonresum2222} remains unchanged under $S \leftrightarrow T$. In fact, the term $-S_1(-S-1)$ can also be fixed from \eqref{eq:2222YMtmp} by imposing this crossing symmetry.

For readers who prefer an approach with more mathematical rigour, we also provide a direct resummation computation of $\widetilde{\mathcal{M}}^{(2)}(S,T)$ in \cref{appx:resum}, using the regularization method mentioned at the beginning of this subsection.

\subsection{Writing Mellin amplitudes as differential operators}

In the above analysis we encounter two closed-form functions $\Phi(S,T)$ and $S_1(-S-1)$ in the Mellin amplitude $\widetilde{\mathcal{M}}^{(2)}(S,T)$. As will be discussed in this and other explicit examples later, together with the function $1$, they form a set of seed functions at the one-loop level in Mellin space. We denote their position-space counterpart via the standard Mellin transformation \eqref{eq:MellinStandardForm} by $\mathcal{W}_4$ and $\mathcal{W}_3$, i.e.,
\begin{align}
	\mathcal{W}_4(z,\bar{z})\equiv\frac{W_4(z,\zb)}{z-\zb} =& \int \frac{\dd S \dd T}{(2\pi i)^2} U^S V^T\, \widetilde{\Gamma}(S,T)\, \Phi(S,T), \label{eq:W4Mellin} \\
	\mathcal{W}_3(z,\bar{z})\equiv\frac{W_3(z,\zb)}{z-\zb} =& \int \frac{\dd S \dd T}{(2\pi i)^2} U^S V^T\, \widetilde{\Gamma}(S,T)\, S_1(-S-1). \label{eq:W3Mellin}
\end{align}
It turns out that the numerator functions $W_4$ and $W_3$ are unital weight SVMPLs in the parity odd sector \footnote{Here we present the expressions of $W_4(1/z,1/\zb)$ and $W_3(1/z,1/\zb)$ as they are shorter in appearance. One can use the function \texttt{ToFibrationBasis} in the Mathematica package \texttt{PolyLogTools} \cite{Duhr:2019tlz} to get the expression of $W_4(z,\zb)$ and $W_3(z,\zb)$, and we also record those expressions in \cref{appx:function}.}
\begin{align}
	W_4\left(\frac{1}{z},\frac{1}{\zb}\right) =&\ G^{\rm sv}_{1101} -G^{\rm sv}_{1011} + 6\zeta_3 G^{\rm sv}_{1}, \\
	W_3\left(\frac{1}{z},\frac{1}{\zb}\right) =&\ G^{\rm sv}_{101} -G^{\rm sv}_{010} + 2 \left[ G_{\zb,z}G_{01,\zb} - G_{\zb,z}G_{10,\zb} - G_{1,\zb}G_{\zb0,z} \right. \nonumber \\
	&\ \left. + G_{0,\zb}G_{\zb1,z} + G_{001,\zb} - G_{101,\zb} - G_{\zb01,z} + G_{\zb10,z}\,\right],
\end{align} 
where we use the abbreviation $G^{\rm sv}_{\vec{a}} \equiv G^{\rm sv}_{\vec{a}}(z,\zb)$ and $G_{\vec{a},z}\equiv G_{\vec{a}}(z)$. The relations \eqref{eq:W4Mellin} and \eqref{eq:W3Mellin} can be verified in the same way as for $\bar{D}_{1111}$, by the contour deformation method described in \cref{sec:tree}. It is worth noting that the function $\mathcal{W}_3(z,\zb)$ also appears in a recent paper \cite{Heslop:2023gzr}  where it is called $\mathfrak{f}_s(z,\zb)$, and is shown to be a $\phi^4$ bubble diagram in ${\rm AdS}_2\times {\rm S}^2$ (and also related to bubble in AdS$_4$, see \cite{Heckelbacher:2022fbx}). We expect $\mathcal{W}_4(z,\zb)$ to be related to the one-loop box diagram in ${\rm AdS}$ space, due to its flat space limit \cite{Alday:2021ajh}, but we leave the explicit computation for the future. 

Given the relations between $\Phi(S,T)$, $S_1(-S-1)$ and $\mathcal{W}_4(z,\zb)$, $\mathcal{W}_3(z,\zb)$, we are ready to work out the differential representation of $\langle 2222 \rangle_{\rm YM}$ at one-loop. The arguments of $\Phi$ in \eqref{eq:gluonresum2222} already suggest us to act $V$ and $V^{-1}$ on $\mathcal{W}_4(z,\zb)$. By rewriting the Mellin transform \eqref{eq:MT2222YM} into the standard form
\begin{align} 
	\mathcal{H}^{(2)}_{st,2222}(U,V) =& \int \frac{\dd S \dd T}{(2\pi i)^2}U^{S+3}V^{T} \widetilde{\mathcal{M}}^{(2)}(S,T)\, \Gamma^2(-S)\Gamma^2(-T)\Gamma^2(3+S+T) \nonumber \\
	=&\ U^3\! \int \frac{\dd S \dd T}{(2\pi i)^2}U^{S}V^{T}\widetilde{\Gamma}(S,T)\ (2+S+T)^2(1+S+T)^2 \nonumber \\
	&\qquad \times\left[ -\frac{2 T^2 }{3 (S+T)}\Phi(S,T-1) + \frac{\left(T^2+T+1\right) }{3 (S+T+1)}\Phi(S,T) \right. \nonumber\\
	&\qquad\qquad \left. +\frac{(T+1)^2 }{3 (S+T+2)}\Phi(S,T+1)- S_1(-S-1)-C \right],
\end{align}
we can read off the differential operators in a similar way as in \cref{sec:tree}, such that
\begin{align} \label{eq:diff2222YM}
	\mathcal{H}^{(2)}_{st,2222}(U,V) =&\ U^3 \left[ -\frac{2}{3}(2 + \DU + \DV)^2(1 + \DU + \DV)^2(\DU + \DV)V \right. \nonumber\\
	&\qquad + \frac{1}{3}(2 + \DU + \DV)^2(1 + \DU + \DV)(1+\DV+\DV^2) \nonumber\\
	&\qquad \left. + \frac{1}{3} (2 + \DU + \DV)(1+\DV)^4 V^{-1}\right] \mathcal{W}_4(z,\zb) \nonumber \\
	&-U^3(2 + \DU + \DV)^2(1 + \DU + \DV)^2 \mathcal{W}_3(z,\zb)  \nonumber \\
	&-C U^3(2 + \DU + \DV)^2(1 + \DU + \DV)^2 \mathcal{W}_2(z,\zb),
\end{align}
where we denote $\mathcal{W}_2(z,\zb)\equiv\bar{D}_{1111}(z,\zb)$. By performing the derivatives, one can verify that this result precisely matches the position space result recorded in \cite{Huang:2023oxf}.

\subsection{The structure of differential representation}\label{sec:diffdiscussion}

The differential representation \eqref{eq:diff2222YM} mimics the tree-level one \eqref{eq:H2222dif}, with two additional functions $\mathcal{W}_4(z,\zb)$ and $\mathcal{W}_3(z,\zb)$. From this result, we can do ``maximal cuts'', which is picking poles in the resumed Mellin amplitude \eqref{eq:gluonresum2222}. By picking poles on $S$ and $T$ simultaneously, we drop the information from $\mathcal{W}_3(z,\zb)$ and $\mathcal{W}_2(z,\zb)$. The remaining residues, $a_{m,n}$, is just the sum of all three rational functions in front of $\Phi$ \footnote{Notice that, the difference between $\Phi$s with different arguments, say $\Phi(S,T+1)$ and $\Phi(S,T)$, is given by $S_1$. To be explicit, $\Phi(S,T+1)-\Phi(S,T) = \frac{S_1(-S-1)-S_1(-T-1)}{T+1}$.}
\begin{equation}
    a_{m,n} = \left. -\frac{2 T^2 }{3 (S+T)} + \frac{\left(T^2+T+1\right) }{3 (S+T+1)} +\frac{(T+1)^2 }{3 (S+T+2)} \right\vert_{S\,\to\, m,\, T\,\to\, n}.
\end{equation}
This is because the residues of $\Phi(S,T)$ on the simultaneous poles on $S$ and $T$ are
\begin{equation}
    \Phi(S,T) \sim \frac{1}{(S-m)(T-n)} + (\text{lower orders}),\quad m,n\geq0.
\end{equation}
One can think the purpose of $\Phi(S,T)$ is to provide a grid of poles with identical residues $1$. Multiplying a rational function $f(S,T)$ in front of $\Phi(S,T)$ will render the residues to be $f(m,n)$, with the side effect of producing additional lower order terms in the Mellin amplitudes. Similarly, the Mellin amplitude of $\mathcal{W}_3(z,\zb)$ has residues
\begin{equation}
    S_1(-S-1) \sim \frac{1}{S-m} + (\text{lower orders}),\quad m\geq0,
\end{equation}
which can be used to generate a row of $S$ poles with residues $1$. Using the linear combination of these two functions and $1$, with the coefficient being rational functions on $S$ and $T$, we can construct nearly any Mellin amplitudes we want with rational residues.

There are still some constraints on the Mellin amplitudes that can be constructed in the differential representation, though. Notice that, the form of the first term appearing in the Mellin amplitude
\begin{equation}\label{eq:poles&zeros}
	-\frac{2 T^2 }{3 (S+T)}\Phi(S,T-1)
\end{equation}
is severely constrained from the view of differential representation. From the operations described in \cref{sec:tree}, we know the denominator $(S+T)$ can only be generated by the multiplication of $U^{a>0}$ and $V^{b>0}$, see rules \eqref{rule:U} and \eqref{rule:V}. In this case, there must also be corresponding {\it double zeros} 
\begin{equation}
    [(-S)_{a}]^2 \equiv S^2(S-1)^2\cdots(S-a+1)^2
\end{equation}
or $[(-T)_{b}]^2$ on the numerators. For \eqref{eq:poles&zeros}, the double zero $T^2$ there confirms our suggestion. In general, any denominators in the standard Mellin amplitudes must correspond to certain double zeros. It serves as simple criteria on determining if it is possible to write a correlator in the differential representation with the three seed functions $\mathcal{W}_{2,3,4}$ mentioned above. Similar conditions also hold for higher loop correlators, and in section \ref{sec:higherloops} we will see they are always satisfied by the leading log part of correlators in ${\rm AdS}_5\times {\rm S}^5$ and ${\rm AdS}_5\times {\rm S}^3$. This is a strong hint on the validity of differential formalism beyond one-loop level. 

Now we turn to the relations between the position space representation and the differential representation. As briefly reviewed in the introduction, in position space the holographic correlators are written as
\begin{equation}\label{eq:positionansatz}
    \sum_{i} \frac{p_i(z,\zb)}{(z-\zb)^n} G_i(z,\zb),
\end{equation}
where $G_i(z,\zb)$ are SVMPLs up to a specific weight (weight 4 for one loop and weight 6 for two loops), and $p_i(z,\zb)$ are some polynomials in $z$ and $\zb$. In bootstrapping holographic correlators in the position space, the polynomials $p_i(z,\zb)$ are usually unknown, and we need to determine them by imposing suitable constraints on the correlator. One obstacle in the bootstrap in position space is that, a correlator should remain finite when taking $z=\zb$ in the Euclidean region,\footnote{Taking $z=\zb$ in the Euclidean region is to take $z$ to be real (since $\zb=z^*$ in the Euclidean region). In this configuration all four operators are placed on a straight line, but it does not lead to any singularities in the correlator.} but this is not manifest in \eqref{eq:positionansatz} because of the denominator $(z-\zb)^{n}$. This finiteness condition has to be resolved by carefully choosing $p_i(z,\zb)$ to induce numerous cancellations between terms with different weights. Fortunately, in the differential representation, the finiteness property becomes transparent. The upshot is that, the SVMPLs $W_{2,3,4}(z,\zb)$ appearing on the numerators of $\mathcal{W}_{2,3,4}(z,\zb)$ are all parity odd, which makes the combination 
\begin{equation}
    \mathcal{W}_{i}(z,\zb) = \frac{W_{i}(z,\zb)}{z-\zb}
\end{equation}
finite when $z=\zb$. Acting differential operators on $\mathcal{W}_i(z,\zb)$ does not change this property, so the differential representation \eqref{eq:diff2222YM} are manifestly finite at $z=\zb$. This also offer us a guiding principle for finding the seed functions, i.e., they should all be unital weight SVMPLs in the parity odd sector, with denominator $z-\zb$.\footnote{In principle, there can also be seed functions with more complicated form. For example, any combination like \eqref{eq:positionansatz} which is finite at $z=\zb$ can be a candidate for the seed functions. This includes the correlator itself, which can be regarded as a seed function with identity operator acting on. Our empirical observation is that, the parity odd SVMPLs seem to be {\it sufficient} for the seed functions in holographic correlators. } This principle greatly reduce the number of functions we need to consider, comparing with those in the position space bootstrap. To be concrete, see table \ref{table:SVMPLs} where we list the number of all SVMPLs with singularities at $0$, $1$, $\infty$ and $z-\zb$ up to weight 6. In general, in the position space bootstrap we may need to use all of these SVMPLs, but the number of possible seed functions (parity odd SVMPLs) is just less than half of that. Especially, there are no parity odd SVMPLs with weight lower than 2.
\begin{table}
\centering
\renewcommand{\arraystretch}{1.2}
    \begin{tabular}{c|ccccccc}
        \hline\hline
        Weight & 0 & 1 & 2 & 3 & 4 & 5 & 6 \\
        \hline
        Odd & 0 & 0 & 1 & 3 & 10 & 28 & 78 \\
        All & 1 & 2 & 5 & 12 & 28 & 69 & 173 \\
        \hline\hline
    \end{tabular}
    \caption{The number of SVMPLs up to weight 6, with singularities at $0$, $1$, $\infty$ and $z-\zb$.}
    \label{table:SVMPLs}
\end{table}

Further detailed structures of the correlator can be extracted by comparing \eqref{eq:positionansatz} with the differential representation \eqref{eq:diff2222YM}. By noticing that taking derivatives on MPLs always lowers their weights, we can conclude that the highest weight terms of a correlator are always generated by acting certain differential operators purely on $(z-\zb)^{-1}$ and then multiplying $W_i(z,\zb)$. For example, in the case of $\mathcal{H}^{(2)}_{st,2222}(U,V)$, the highest weight part is given by
\begin{equation}
    W_4(z,\zb)\, \mathcal{D}_{\mathcal{W}_4} \left(\frac{1}{z-\zb}\right),
\end{equation}
where $\mathcal{D}_{\mathcal{W}_4}$ is the differential operator corresponds to $\mathcal{W}_{4}(z,\zb)$ in the differential representation \eqref{eq:diff2222YM}. As a consequence, the highest weight $G_i(z,\zb)$ in \eqref{eq:positionansatz} always falls into the parity odd sector (as one can easily check for all the correlators in ${\rm AdS}_5 \times {\rm S}^5$ and ${\rm AdS}_5 \times {\rm S}^3$ in literature), and we can read off the highest weight seed functions directly from the position space representation. Moreover, under suitable assumptions, we can use the rational functions in \eqref{eq:positionansatz} to reconstruct the operators in the differential representation.

There are also some comments on the seed functions. By now, we see these functions $\mathcal{W}_{2,3,4}(z,\zb)$ are unital weight in position space, but they are also unital weight functions ($1$, $S_1$ and $\Phi$) in Mellin space. Moreover, their weights in Mellin space are always lower by 2 comparing with the corresponding MPLs. This missing 2 should be accounted for by the $\Gamma$ factor $\widetilde{\Gamma}(S,T)$ in the Mellin transformation. One may wonder if the other two independent weight 2 harmonic sums, $S_{2}(-S-1)$ and $S_{1,1}(-S-1)$ play any roles in the holographic correlators. In position space, these two functions correspond to 
\begin{align}
    \mathcal{W}_{4,1}(z,\bar{z})\equiv\frac{W_{4,1}(z,\zb)}{z-\zb} =& \int \frac{\dd S \dd T}{(2\pi i)^2} U^S V^T\, \widetilde{\Gamma}(S,T)\, S_{2}(-S-1), \label{eq:W41Mellin} \\
	\mathcal{W}_{4,2}(z,\bar{z})\equiv\frac{W_{4,2}(z,\zb)}{z-\zb} =& \int \frac{\dd S \dd T}{(2\pi i)^2} U^S V^T\, \widetilde{\Gamma}(S,T)\, S_{1,1}(-S-1), \label{eq:W42Mellin}
\end{align}
with
\begin{align}
	W_{4,1}\left(\frac{1}{z},\frac{1}{\zb}\right) =&\ G^{\rm sv}_{1010} -G^{\rm sv}_{0101} + \zeta_2 \left(G_{10}^{\rm sv}-G_{01}^{\rm sv}\right)- 4\zeta_3 G^{\rm sv}_{1}, \\
	W_{4,2}\left(\frac{1}{z},\frac{1}{\zb}\right) =&\ 
	G_{1,z} G_{001,\zb}-G_{1,z} G_{101,\zb}+G_{01,\zb} G_{0\zb ,z}-G_{10,\zb} G_{0\zb ,z}+G_{01,\zb} G_{1\zb ,z} \nonumber\\
	& -G_{10,\zb} G_{1\zb ,z} +2 G_{01,\zb} G_{\zb0,z} -2 G_{10,\zb} G_{\zb1,z}-4 G_{01,\zb} G_{\zb\zb ,z}+4 G_{10,\zb} G_{\zb\zb ,z}\nonumber\\
	& +G_{0,z} G_{001,\zb} -4 G_{001,\zb} G_{\zb ,z}  +2 G_{010,\zb} G_{\zb ,z}-G_{1,\zb} G_{0\zb0,z}+G_{0,\zb} G_{0\zb1,z} \nonumber\\
	& -G_{0,z}G_{101,\zb}+2 G_{101,\zb} G_{\zb ,z}-G_{1,\zb} G_{1\zb0,z} +G_{0,\zb} G_{1\zb1,z}+2 G_{0,\zb} G_{\zb01,z}\nonumber\\
	& -2 G_{1,\zb} G_{\zb10,z}+4 G_{1,\zb} G_{\zb\zb0,z}-4 G_{0,\zb} G_{\zb\zb1,z}-G_{0\zb01,z}  +G_{0\zb10,z}-G_{1\zb01,z} \nonumber\\
	& +G_{1\zb10,z}+2 G_{\zb010,z}-2G_{\zb101,z}+4 G_{\zb\zb01,z}-4 G_{\zb\zb10,z} -3 G_{0001,\zb}+G_{0010,\zb}\nonumber\\
	& +2 G_{0101,\zb}+G_{1001,\zb}-G_{1010,\zb}  -4 \zeta_3 G_{1,z} +4 \zeta_3 G_{1,\zb} + \zeta_2 \left(G_{10}^{\rm sv}-G_{01}^{\rm sv}\right) .
\end{align}
The function $\mathcal{W}_{4,1}$ appears in the leading log part of two-loop correlators, as we will soon see in section \ref{sec:higherloops}. We believe that $\mathcal{W}_{4,2}$ will appear in the stringy corrections at the two-loop level. One can also check that, $W_4$, $W_{4,1}$ and $W_{4,2}$ form a complete basis (upon crossing symmetry) of weight 4 SVMPLs in the parity odd sector, while their Mellin counterparts $\Phi$, $S_1$ and $S_{1,1}$ form a complete basis of weight 2 harmonic sums.

\section{More examples at one loop}\label{sec:oneloop}

In this section, we provide more explicit examples to show the benefits and versatility of the differential representation. The examples include 
\begin{itemize}
    \item $\langle 3333 \rangle_{\mathrm{YM}}$ in ${\rm AdS}_5\times {\rm S}^3$. This is a one-loop correlator with higher KK-modes. We resum the Mellin amplitude for $\langle3333\rangle_{\mathrm{YM}}$ and use the differential representation to provide the position space results of it. 
    \item $\langle 2222 \rangle_{\mathrm{GR}}$ in ${\rm AdS}_5\times {\rm S}^5$. The position space result of $\langle 2222\rangle_{\mathrm{GR}}$ is obtained in \cite{Aprile:2017bgs} and in \cite{Alday:2018kkw} there is the Mellin amplitude. Using the differential representation, we provide a direct relation for the result on both sides. It is worth noting that, in this case we only need to use the seed function $\mathcal{W}_4$ (up to the possible ambiguity containing $\mathcal{W}_2$).
    \item One-loop stringy corrections of $\langle 2222 \rangle_{\mathrm{GR}}$ in ${\rm AdS}_5\times {\rm S}^5$. These stringy corrections \cite{Alday:2018kkw,Chester:2019pvm,Drummond:2019hel} contain the seed functions $\mathcal{W}_2$ and $\mathcal{W}_3$, but there is no $\mathcal{W}_4$ in them. We revisit the one-loop stringy Mellin amplitude in the first few orders and reorganize them into differential representation. 
\end{itemize}

\subsection{$\langle 3333 \rangle_\mathrm{YM}$ in ${\rm AdS}_5\times {\rm S}^3$}

In this subsection, we still focus on the correlators in ${\rm AdS}_5\times {\rm S}^3$ at one-loop level, but extend the results to higher KK-modes $\langle 3333 \rangle_\mathrm{YM}$. Similar to $\langle 2222 \rangle_\mathrm{YM}$, we can decompose the one-loop correlators in different channels according to their color structures 
\begin{equation}
	\widetilde{\mathcal{M}}^{\{I_i\},(2)}_{\mathrm{YM},3333} = \dst\, \widetilde{\mathcal{M}}^{(2)}_{st}(S,T;\alpha,\beta) + \dsu\, \widetilde{\mathcal{M}}^{(2)}_{su}(S,\tilde{U};\alpha,\beta) + \dtu\, \widetilde{\mathcal{M}}^{(2)}_{tu}(T,\tilde{U};\alpha,\beta)\, .
\end{equation}
The Mellin amplitudes in each channel are related by crossing symmetry, so let us focus on $\widetilde{\mathcal{M}}^{(2)}_{st}(S,T;\alpha,\beta)$. This amplitude only involves simultaneous poles \cite{Huang:2023ppy},
\begin{equation}\label{eq:gluon3333oneloop}
	\widetilde{\mathcal{M}}^{(2)}_{st}(S,T;\alpha,\beta) = \sum_{m,n=-1}^{\infty} \frac{b^{st}_{m,n}}{(S-m)(T-n)} \, .
\end{equation}
The complete expression of the residues $b_{m,n}$ is 
\begin{equation}\label{eq:bulkres3333}
    b^{st}_{m,n}= \begin{cases}\displaystyle{\frac{F^{0,0}_{m,n}}{3(m+n+1)_3}\!-\!\frac{F^{1,0}_{m,n}\alpha^{-1}}{3(m+n+1)_3}\!-\!\frac{F^{0,1}_{m,n}\beta^{-1}}{3(m+n+1)_3}\!+\!\frac{F^{1,1}_{m,n}(\alpha\beta)^{-1}}{3(m+n)_4}, \quad m,n\geq0 \, ,}   \\[14pt] 
    \displaystyle{\frac{3 m+5}{2 (m+1)_2}\!-\!\frac{(3 m+4)\alpha^{-1}}{2  (m+1)_2}\!-\!\frac{(3 m+5)\beta^{-1}}{2(m+1)_2} \!+\!\frac{(3 m+4)(\alpha\beta)^{-1}}{2  (m+1)_2}}, \quad  n=-1,\ m\geq0 \, , \\[14pt] 
    \displaystyle{\frac{3 n+5}{2 (n+1)_2}-\frac{\alpha^{-1}}{2  (n+1)_2}} ,  \quad   m=-1,\ n\geq0 \, , \\[14pt]
    \ 0, \quad m=n=-1.
    \end{cases}
\end{equation}
Here $\alpha$, $\beta$ are the variables of R-symmetry structures. The polynomials $F^{i,j}_{m,n}$ reads
\begin{subequations}
\begin{align}
    F^{0,0}_{m,n}=&\ 20 m^2 n+22 m^2+20 m n^2+96 m n+73 m+22 n^2+73 n+40 \, ,\\
    F^{1,0}_{m,n}=&\ 10 m^2 n+14 m^2+10 m n^2+48 m n+41 m+8 n^2+32 n+20 \, , \\
    F^{0,1}_{m,n}=&\ 10 m^2 n+14 m^2+10 m n^2+48 m n+44 m+8 n^2+29 n+20 \, , \\
    F^{1,1}_{m,n}=&\ 124 m^2 n+118 m n^2+76 m^2+64 n^2+167 m n +22 m^3+16 n^3 \nonumber\\
    &+40 m^2 n^2+52 m+52 n+20 m^3 n+20 m n^3\, . 
\end{align}
\end{subequations}
Next, we proceed with the summation of this residues. When $m,n\geq0$, we rewrite the residues $b^{st}_{m,n}$ as
\begin{align}
     b^{st}_{m,n}=&\quad \left(-\frac{32 n^2+32 n+11}{6 (m+n+1)}-\frac{(m-1)-(8 n^2+15 n+9)}{2 (m+n+2)}+\frac{(n+2) (8 n+11)+3(m+2)}{6 (m+n+3)} \right)   \nonumber\\
     +& \frac{1}{\alpha} \left( \frac{16 n^2+19 n+7}{6 (m+n+1)}-\frac{(4 n^2+7 n+4)+(m+2)}{2 (m+n+2)}+\frac{3(m+2)-2(n+2) (2 n+5)}{6 (m+n+3)}\right) \nonumber \\
     +& \frac{1}{\beta} \left(\frac{16 n^2+25 n+10}{6 (m+n+1)} -\frac{2 n^2+5 n+4}{m+n+2} -\frac{4 n^2+15 n+14}{6 (m+n+3)}  \right) \nonumber \\
     +& \frac{1}{\alpha\beta}\left(-\frac{3 n^2}{2 (m+n)} -\frac{17 n^2+23 n+2}{6 (m+n+1)}+\frac{7 n^2+14 n+8}{2 (m+n+2)}+\frac{(n+2)(5n+11)}{6 (m+n+3)} \right) \, .
\end{align}
We can perform the complete resummation for each of the R-symmetry channels individually. We point out that there is a double zero $-3n^2/(2(m+n))$ in the residues, which allows us to carry out the differential representation. Since we have done the partial fractions, the results of their summation can be derived directly
\begin{align}
    \mathcal{F}^{0,0}_{S,T}=& -\frac{32 T^2+32 T+11}{6 (S+T+1)} \Phi(S,T) -\frac{S-1}{2(S+T+2)} \Phi(S+1,T)+\frac{8 T^2+15 T+9}{2 (S+T+2)}  \Phi(S,T+1) 
     \nonumber \\
    & +\frac{S+2 }{2 (S+T+3)}\Phi(S+2,T)+\frac{(T+2) (8 T+11) }{6 (S+T+3)} \Phi(S,T+2)- \frac{20}{3}S_1(-S-1)\, , \\
    \mathcal{F}^{1,0}_{S,T}=&\frac{16 T^2+19 T+7}{6 (S+T+1)}\Phi (S,T) -\frac{S+2}{2 (S+T+2)}\Phi (S+1,T)-\frac{4 T^2+7T+4}{2 (S+T+2)} \Phi (S,T+1)\nonumber \\
    &+\frac{S+2}{2 (S+T+3)}\Phi (S+2,T)-\frac{(T+2) (2 T+5)}{3 (S+T+3)} \Phi (S,T+2) + \frac{10}{3}S_1(-S-1)\, ,\\
    \mathcal{F}^{0,1}_{S,T}=&\frac{ 16 T^2+25 T+10 }{6 (S+T+1)} \Phi (S,T)-\frac{2 T^2+5 T+4}{S+T+2}\Phi (S,T+1) -\frac{4 T^2+15 T+14}{6 (S+T+3)} \Phi (S,T+2)  \nonumber\\
    &+\frac{10}{3}S_1(-S-1)\, ,  \\[7pt]
    \mathcal{F}^{1,1}_{S,T}=&-\frac{3 T^2 }{2 (S+T)}\Phi (S,T-1)-\frac{17 T^2+23 T+2 }{6 (S+T+1)}\Phi (S,T)+\frac{ 7 T^2+14 T+8 }{2 (S+T+2)} \Phi (S,T+1) \nonumber \\ 
    &+\frac{ (T+2)(5T+11) }{6 (S+T+3)}\Phi (S,T+2) -\frac{20}{3}S_1(-S-1)\, .
\end{align}
Remarkably, these analytic functions $\mathcal{F}^{i,j}_{S,T}$ also precisely predicts the the residues \eqref{eq:bulkres3333} in the cases of $m=-1$ or $n=-1$. The final Mellin amplitude $\widetilde{\mathcal{M}}^{(2)}_{st}(S,T;\alpha,\beta)$ yields
\begin{align}\label{eq:Mellin3333}
    \widetilde{\mathcal{M}}^{(2)}_{st}=\mathcal{F}^{0,0}_{S,T}+\frac{1}{\alpha}\mathcal{F}^{1,0}_{S,T}+\frac{1}{\beta} \mathcal{F}^{0,1}_{S,T}+\frac{1}{\alpha \beta}\mathcal{F}^{1,1}_{S,T} + C_1\left(1-\frac{1}{2 \alpha}-\frac{1}{2 \beta}\right)+C_2\left(\frac{1}{\alpha \beta}\right),
\end{align}
where the $C_1$ and $C_2$ parts correspond to the UV divergence in AdS. Readers can check that \eqref{eq:Mellin3333} is invariant under exchanging operators $1\leftrightarrow3$. 

In order to convert \eqref{eq:Mellin3333} into the differential representation, it is convenient to express the Mellin amplitude in its standard form
\begin{align} 
	U^{-4}\mathcal{H}^{(2)}_{st}(U,V;\alpha,\beta) =& \int \frac{\dd S \dd T}{(2\pi i)^2}U^{S}V^{T}\widetilde{\Gamma}(S,T) ((1+S+T)_3 )^2 \widetilde{\mathcal{M}}^{(2)}_{st}(S,T;\alpha,\beta) \,  .
\end{align}
By transforming all rational functions of $S$ and $T$ into differential operators, we present the position space result as
\begin{align} 
	\mathcal{H}^{(2)}_{st}(U,V;\alpha,\beta) =& U^4 \left(\mathcal{D}^{0,0} +\frac{1}{ \alpha}\mathcal{D}^{1,0} +\frac{1}{\beta}\mathcal{D}^{0,1} + \frac{1}{\alpha \beta}\mathcal{D}^{1,1} \right)  \mathcal{W}_4(z,\zb) \nonumber \\
	&-\frac{10 (2 \alpha  \beta -\alpha -\beta +2)}{3 \alpha  \beta }U^4  \big[(\DU+\DV+1)_3\big]^2 \mathcal{W}_3(z,\zb)  \nonumber \\
	&+ \frac{C_1(2\alpha\beta-\alpha-\beta)+2C_2}{2\alpha\beta}U^4  \big[(\DU+\DV+1)_3\big]^2 \mathcal{W}_2(z,\zb) \, ,
\end{align}
where the differential operators $\mathcal{D}^{i,j}$ are
\begin{align}
    \mathcal{D}^{0,0} =
    & -\frac{1}{2} ( \DU-1)  (1 +  \DU)^2  \Delta_{2,3} U^{-1} + \frac{1}{2} ( \DV+1)^2  \left(8  \DV^2+15  \DV+9\right)  \Delta_{2,3} V^{-1}  \nonumber \\
    &+ \frac{1}{2} ( \DU+1)^2 ( \DU+2)^3 \Delta_{3,3} U^{-2}  + \frac{1}{2} ( \DV+1)^2 ( \DV+2)^3 (8 \DV+11) \Delta_{3,3} V^{-2} \nonumber \\
    &-\frac{1}{6} \left(32  \DV^2+32  \DV+11\right)\Delta_{1,3} \, , \\
    \mathcal{D}^{1,0}=&- \frac{1}{2}( \DU+1)^2 ( \DU+2) \Delta_{2,3} U^{-1} - \frac{1}{2} ( \DV+1)^2 \left(4  \DV^2+7  \DV+4\right) \Delta_{2,3}  V^{-1}  \nonumber \\
    &+ \frac{1}{2} ( \DU+1)^2 ( \DU+2)^3 \Delta_{3,3} U^{-2}     -\frac{1}{3} (  \DV+1)^2 ( \DV+2)^3 (2  \DV +5) \Delta_{3,3} V^{-2} \nonumber \\
    & +\frac{1}{6} \left(16  \DV^2+19  \DV+7\right) \Delta_{1,3} \, ,  \\
    \mathcal{D}^{0,1}=& \frac{1}{6} \left(16 \DV^2+25 \DV+10\right)  \Delta_{1,3} - ( \DV+1)^2  (2  \DV^2+5  \DV+4) \Delta_{2,3} V^{-1}  \nonumber \\
    &- \frac{1}{6} ( \DV+1)^2 ( \DV+2)^3 (4  \DV+7)\Delta_{3,3} V^{-2}  \, , \\
    \mathcal{D}^{1,1}=& -\frac{3}{2} \Delta_{0,3} V  - \frac{1}{6}  \left(17 \DV^2+23 \DV+2\right) \Delta_{1,3}  + \frac{1}{2} ( \DV+1)^2 \left(7  \DV^2+14  \DV+8\right)  \Delta_{2,3} V^{-1}  \nonumber \\
    &  + \frac{1}{6} ( \DV+1)^2 ( \DV+2)^3 (5  \DV+11) \Delta_{3,3} V^{-2} \, .
\end{align}
Here we introduce the notation $\Delta_{i,j}$
\begin{align}
    \Delta_{i,i} =& (i+\DU+\DV)\, ,\\
    \Delta_{i,j} =& (i+\DU+\DV)(1+i+\DU+\DV)^2\cdots(j+\DU+\DV)^2 \,,
\end{align}
for convenience. Notice that in $\Delta_{i,j}$ the power of the first term $(i+\DU+\DV)$ is 1.

\subsection{$\langle 2222 \rangle_\mathrm{GR}$ in ${\rm AdS}_5\times {\rm S}^5$}
In this subsection, we turn to type IIB supergravity theory in ${\rm AdS}_5\times {\rm S}^5$ and focus on the lowest KK mode $\langle 2222\rangle_\mathrm{GR}$. Supergraviton amplitudes are very analogous to supergluon cases so we can apply similar analysis directly. The reduced correlator $\mathcal{H}^{(2)}_{\mathrm{GR},\, 2222}$ is related to the Mellin amplitude through
\begin{align}
    \mathcal{H}_{\mathrm{GR},\,2222}^{(2)}=\int\frac{\mathrm{d}S\mathrm{d}T}{(2\pi i)^2}U^{S+4}V^{T} \widetilde{\mathcal{M}}^{(2)}_{\mathrm{GR},\,2222}(S,T)\,\widetilde{\Gamma}_{2222}(S,T)\, ,
\end{align}
where
\begin{equation}
    \widetilde{\mathcal{M}}^{(2)}_{\mathrm{GR},\,2222}(S,T)=\widetilde{\mathcal{M}}^{(2)}(S,T)+\widetilde{\mathcal{M}}^{(2)}(\tilde{U},T)+\widetilde{\mathcal{M}}^{(2)}(S,\tilde{U}).
\end{equation}
The structure of this Mellin amplitude only contains simultaneous poles
\begin{equation}
    \widetilde{\mathcal{M}}^{(2)}(S,T)=\sum_{m,n=0}^{\infty} \frac{c_{m,n}}{(S-m)(T-n)}.
\end{equation}
In \cite{Alday:2018kkw} the author derives the expression of the residues 
\begin{equation}
    c_{m,n}=\frac{16}{5 (m+n-1)_5} (F_{m,n}+F_{n,m}),
\end{equation}
where $F_{m,n}$ is a polynomial of $m$ and $n$
\begin{align} 
   F_{m,n} =& \quad  2 (m-1) m (n+1) (n+2) (m+n+2) (m+n+3) \nonumber\\
           &+ (m+1) (m+2) (n+1) (n+2) (m+n-1) (m+n) \nonumber\\
           &+4 m (m+1) n (n+1)(m+n+2) (m+n+3)\nonumber\\
           &+8 m (m+1) (n+1) (n+2) (m+n-1) (m+n+3).
\end{align}

The resummation of $\langle 2222\rangle_\mathrm{GR}$ is also similar to the supergluon cases. According to crossing symmetry, we rewrite the residues $c_{m,n}$ in the symmetric form
\begin{equation}
    c_{m,n}=f_{m,n}+f_{n,m} \, ,
\end{equation}
where
\begin{align}
    f_{m,n}=&\ \frac{48 (m-1)^2 m^2}{5 (m+n-1)}-\frac{8 \left(4 m^2+19 m+20\right) m^2}{5 (m+n)} +\frac{16 (m+1)^2 \left(3 m^2-m+1\right)}{5 (m+n+2)} \nonumber \\
    &+\frac{8 (m+1)^2 (m+2)^2}{5 (m+n+3)} -\frac{4 \left(18 m^4-30 m^3-64 m^2+5 m n-44 m-22\right)}{5 (m+n+1)} \, .
\end{align}
By now the resummation becomes straightforward. After utilizing the function $S_1$ to cancel all of the single poles, the Mellin amplitudes yields \footnote{See also the appendix B in \cite{Alday:2021vfb}.}
\begin{align}\label{eq:graMst}
    \widetilde{\mathcal{M}}^{(2)}(S,T)=& \ \frac{8}{5}\frac{(S+1)^2 (S+2)^2}{ S+T+3} \Phi (S+2,T) +\frac{16}{5}\frac{ \left(3 S^2-S+1\right)(S+1)^2 }{ S+T+2}\Phi (S+1,T)  \nonumber \\
    &-\frac{4}{5}\frac{ \left(18 S^4-30 S^3-64 S^2+5 S T-44 S-22\right)}{S+T+1} \Phi (S,T)\nonumber\\
    & -\frac{8}{5}\frac{\left(4 S^2+19 S+20\right) S^2 }{ S+T}\Phi (S-1,T) +\frac{48}{5}\frac{ (S-1)^2 S^2 }{ S+T-1} \Phi (S-2,T) \nonumber   \\
    &-16 (S+2 T+4) S_1(-S-1)  + C_0 +(S\leftrightarrow T).
\end{align}
First of all, let us check the preceding rational functions of $\Phi$. For the functions $\Phi(S-2,T)$ and $\Phi(S-1,T)$, there exist corresponding double zeros in their coefficients, allowing us to write down the differential operators. The entire $\widetilde{\mathcal{M}}^{(2)}_{\mathrm{GR},2222}$ is the sum over the additional contribution of \eqref{eq:graMst} under crossing. Notably, in the final result the function $S_1$ automatically cancels out by substituting $S+T+\tilde{U}=-4$.

The absence of function $S_1$ indicates that the reduced correlator ${\mathcal{H}}^{(2)}_{\mathrm{GR},2222}$ only involves two kinds of functions, $\mathcal{W}_{2}$ and $\mathcal{W}_{4}$. Indeed, the complete expression the reduced correlator can be written as
\begin{align}
    \mathcal{H}_{\mathrm{GR},\,2222}^{(2)} =(\, \mathcal{I} (U,V) + (\text{crossing})\,  )  + C_0^{'} (\DU+\DV+1) \Delta_{1,3} \mathcal{W}_2(z,\zb)
\end{align}
where we introduce
\begin{align}\label{eq:gra2222H}
    \mathcal{I} (U,V)=
     U^4 & \left[\frac{8}{5} (\DU+1)^4 (\DU+2)^4 \Delta_{3,3} U^{-2}+\frac{16}{5}  \left(3 \DU^2-\DU+1\right)(\DU+1)^4 \Delta_{2,3} U^{-1}  \right. \nonumber \\
    & -\frac{4}{5}(18 \DU^4-30 \DU^3-64 \DU^2+5 \DU \DV-44 \DU-22) \Delta_{1,3}\nonumber  \\
    &\left. -\frac{8}{5} \left(4 \DU^2+19 \DU+20\right) \Delta_{0,3} U +\frac{48}{5}\Delta_{-1,3} U^2 \right] \mathcal{W}_{4}(z,\zb)  \, .
\end{align}
Here we provide a comment on \eqref{eq:gra2222H}. The crossing term include the other five crossing symmetric parts of the function $\mathcal{I}(U,V)$, which are controlled by the function $\mathcal{W}_4$. The remaining contribution involving function $\mathcal{W}_2$ corresponds to the ambiguity from the UV divergence. Unlike the case of supergluons, there is no color structures to distinguish different channels here, and as mentioned earlier the function $\mathcal{W}_3$ vanishes under crossing. Therefore, we conclude that for one-loop supergravity, we do not need the function $\mathcal{W}_3$, which may imply that we do not need the contribution from bubble diagrams.

\subsection{One-loop stringy corrections of $\langle 2222 \rangle_\mathrm{GR}$ in ${\rm AdS}_5\times {\rm S}^5$}\label{sec:stringy}

We can also apply our representation to stringy corrections. Following the works \cite{Alday:2018kkw,Chester:2019pvm,Drummond:2019hel}, we focus on the simplest supergravity case $\langle2222\rangle_\mathrm{GR}$. We first take the large $N$ limit with large but fixed 't Hooft coupling $\lambda =g^2_{\mathrm{YM}} N$. The reduced amplitude $\widetilde{\mathcal{M}}_{\mathrm{GR}}$ takes the form\footnote{At order $a^2$, we follow the conventions in \cite{Drummond:2019hel} and omit all the contact terms which are polynomials on $S$ and $T$. These terms include the super-leading term $\lambda^{\frac{1}{2}} \widetilde{\mathcal{M}}^{(2,-1)}$ and an additional term $\lambda^{-1} \widetilde{\mathcal{M}}^{(2,2)}$. We thank Hynek Paul for pointing out this. }
\begin{align}
    \widetilde{\mathcal{M}}_{\mathrm{GR}}=\quad  a &\left( \widetilde{\mathcal{M}}^{(1,0)}+\lambda^{-\frac{3}{2}}\widetilde{\mathcal{M}}^{(1,3)}+\lambda^{-\frac{5}{2}}\widetilde{\mathcal{M}}^{(1,5)}+\cdots\right) \nonumber \\
    + \; a^2 &\left(\widetilde{\mathcal{M}}^{(2,0)}+\lambda^{-\frac{3}{2}}\widetilde{\mathcal{M}}^{(2,3)}+\lambda^{-\frac{5}{2}}\widetilde{\mathcal{M}}^{(2,5)}+\cdots \right)+\mathcal{O}(a^3).
\end{align}
where $a \equiv a_C/4 = 1/(N^2 - 1)$ is the large $N$ expansion parameter. In this subsection we introduce the notation $\widetilde{\mathcal{M}}^{(m,n)}$ as \cite{Drummond:2019hel} did, where $m$ denotes the order of $a$ and $n$ represents the order of $\lambda^{-\frac{1}{2}}$.

Let us briefly review the tree-level  Mellin amplitude. We have discussed in detail the well-known tree-level supergravity result $\widetilde{\mathcal{M}}^{(1,0)}$ in section \ref{sec:tree}, which can be organized as a differential operator acting on $\bar{D}_{1111}(U,V)$. The higher-order tree-level stringy correction can be interpreted from the perspective of the effective action of supergravity in ${\rm AdS}_5\times {\rm S}^5$. These contributions are only polynomials of $\sigma_2=s^2+t^2+\tilde{u}^2$ and $\sigma_3=s^3+t^3+\tilde{u}^3$ in the Mellin space, which can be regarded as contact terms in AdS. For fixed stringy correction $\lambda^{-\frac{k}{2}}$, the  Mellin amplitude $\widetilde{\mathcal{M}}^{(1,k)}$ is supported by finite spin, hence the power of these polynomials has a truncation,
\begin{equation}
    \sigma_2^{p}\sigma_3^q+\text{subleading terms},
\end{equation}
where $3+2p+3q=k$. We can directly use \eqref{rule:VdV} and \eqref{rule:VVdVV} to ``translate'' them into differentiations on $\bar{D}_{1111}(U,V)$.
We choose $\widetilde{\mathcal{M}}^{(1,3)}$ and $\widetilde{\mathcal{M}}^{(1,5)}$ as examples. Their expression in Mellin space are \footnote{Notice that in our convention we have $s=2S+4$, $t=2T+4$ and $\tilde{u}=2\tilde{U}+4$.}
\begin{align}
    \widetilde{\mathcal{M}}^{(1,3)}&=7680 \zeta_3\, ,\\
    \widetilde{\mathcal{M}}^{(1,5)}&=40320\zeta_5 \left(8 S^2+8 S T+32 S+8 T^2+32 T+45 \right).
\end{align}
Their position space results read
\begin{align}
    \mathcal{H}^{(1,3)}=& 7680\zeta_3  U^4 \,  \mathcal{J}(U,V)\, , \\
    \mathcal{H}^{(1,5)}=& 40320\zeta_5 U^4 \left(8\DU^2+ 8 \DU\DV+32\DU+ 8 \DV^2+32\DV+45\right) \mathcal{J}(U,V)\, ,
\end{align}
where
\begin{align}
     \mathcal{J}(U,V)&=\int\frac{\mathrm{d}S\mathrm{d}T}{(2\pi i)^2}U^{S}V^T\Gamma(-S)^2\Gamma(-T)^2\Gamma(4+S+T)^2 \nonumber \\
     &=\left(1+\DU +\DV\right)^2\left(2+\DU +\DV\right)^2\left(3+\DU +\DV\right)^2 \mathcal{W}_2(z,\zb)\, .
\end{align}
$\mathcal{H}^{(m,n)}$ and $\widetilde{\mathcal{M}}^{(m,n)}$ are related by the Mellin transform \eqref{eq:MtoHnew}, where there is an overall factor $4U^2$ comparing with \cite{Drummond:2019hel}. As we can see, in the case of tree-level stringy correction, the results in the Mellin space can be quickly and directly converted into the position space. 

Next, we will continue to use this method in the one-loop  stringy correction. Following \cite{Drummond:2019hel}, we suppress all the contact terms, which are just polynomials in $\sigma_2$ and $\sigma_3$. And the Mellin amplitudes of one-loop stringy corrections have the structure
\begin{equation}
    \widetilde{\mathcal{M}}^{(2,k)}(S,T)=\sum_{m+n=k} f^{m|n}(S,T)S_1(-S-1)+ (S \leftrightarrow \tilde{U})+ (T \leftrightarrow \tilde{U} )\, .
\end{equation}
Here $f^{m|n}(S,T)$ are polynomials in $S$ and $T$ that satisfy
\begin{equation}
    f^{m|n}(S,T)=f^{m|n}(S,\tilde{U})\, .
\end{equation}
This expression only involves one function $S_1$ and its crossed versions. We already know that the polynomials $f^{m|n}$ can be directly translated into differential operators. Therefore, we can obtain the position space results from the known Mellin amplitudes straightforwardly. For instance, consider
\begin{align}
    f^{0|3}(S,T)=&-1024 \; \zeta (3) \left(63 S^4+182 S^3+273 S^2+202 S+60\right)  \, , \\
    f^{3|3}(S,T)=&-\frac{276480}{7}\zeta_3^2\left( 924 S^7+1309 S^6+5229 S^5 \right. \nonumber \\
    & \quad  \left. +5425 S^4 + 6391 S^3 +3766 S^2+1436 S+138588 \right)\, ,\\
    f^{0|5}(S,T)=&-256\zeta_5\left(21780 S^6 +S^5 (990 T+53235) +S^4 \left(990 T^2+8100 T+137601\right) \right.\nonumber \\
    & \quad +S^3 \left(4140 T^2+24426 T+184063\right) + S^2 \left(7866 T^2+38628 T+173751\right) \nonumber\\
    & \quad \left.+S \left(7164 T^2+31176 T+94670\right)+2520 T^2+10080 T+23100\right)\, .
\end{align}
Using the correspondence 
\begin{align}
    (\DU)^a \mathcal{H}(U,V) \quad   &\Leftrightarrow \quad {S}^a\widetilde{\mathcal{M}}(S,T) \, , \\
    (\DV)^b \mathcal{H}(U,V) \quad   &\Leftrightarrow \quad {T}^a\widetilde{\mathcal{M}}(S,T) \, ,
\end{align}
one can derive the position space results in a direct manner
\begin{equation}
    \mathcal{H}^{(2,k)}=\sum_{m+n=k} U^4 f^{m|n}(\DU,\DV)\mathcal{K}(U,V)+ (\text{crossing})\, ,
\end{equation}
where we define
\begin{equation}
     \mathcal{K}(U,V) =\left(1+\DU +\DV\right)^2\left(2+\DU +\DV\right)^2\left(3+\DU +\DV\right)^2 \mathcal{W}_3(z,\bar{z}) \, .
\end{equation}
In addition, we can also extend this method to stringy correction in ${\rm AdS}_5\times {\rm S}^3$ \cite{Paul:2023zyr}. We conclude that the one-loop stringy corrections in these two backgrounds can be both expressed as the polynomials of $\DU$ and $\DV$ acting on $\mathcal{W}_3$ in position space.

\section{Towards higher-loop level}\label{sec:higherloops}

The correlators we extensively discussed above have already shown the applicability of differential representation at the one-loop level. In this section, we present strong evidence that this representation continues to work for correlators in higher loops. To demonstrate, we will work on $\langle 2222 \rangle_\mathrm{GR}$ at the two- and three-loop level. 

Very little are known about the holographic correlators beyond one-loop level, except for two-loop $\langle 2222 \rangle_\mathrm{GR/YM}$, which were computed in position space in \cite{Huang:2021xws,Drummond:2022dxw,Huang:2023oxf}. Another exception is the leading log part of a correlator, for which there are all-loop results for arbitrary KK modes based on the hidden conformal symmetry. This beautiful result is first derived in ${\rm AdS}_5 \times {\rm S}^5$ \cite{Caron-Huot:2018kta}, which states that the leading log part of $\mathcal{H}^{(k)}_\mathrm{GR}$ is given by
\begin{equation}\label{eq:llk}
    \mathcal{H}^{(k)}_{\mathrm{GR},\{p_i\}}\Big|_{\log^k U} = \big[ \Delta^{(8)} \big]^{k-1} \mathcal{D}_{\{p_i\}} \mathcal{D}_{(3)} h^{(k)}(z)\,,
\end{equation}
where $\Delta^{(8)}$ and $\mathcal{D}_{\{p_i\}}$ are some differential operators depending on $p_i$, and for $p_i=2$ they take the following form,\footnote{The general definition of $\Delta^{(8)}$ and $\mathcal{D}_{\{p_i\}}$ can be found in the section 5 in \cite{Caron-Huot:2018kta}.} 
\begin{align}
    \mathcal{D}_{2222} =&\, 1\; ,\\
    \Delta^{(8)} =&\, \big[ \Delta^{(2)} \big]^2 U^{-4}V^2 \big[ \Delta^{(2)} \big]^2\; , \\
    \Delta^{(2)} =&\, U^3\partial_U^2 + U^2V \partial_V^2 + U^2(-1+U+V)\partial_U\partial_V + U(1+U-V)\partial_V\, .
\end{align}
Moreover, $\mathcal{D}_{(3)}$ is a third-order differential operator
\begin{align}
     \mathcal{D}_{(3)} = &\left[ \left(\frac{z \zb}{\zb - z}\right)^7  + \left(\frac{z \zb}{\zb - z}\right)^6 \frac{z^2}{2}\partial_z + \left(\frac{z \zb}{\zb - z}\right)^5 \frac{z^3}{10}\partial^2_z z + \left(\frac{z \zb}{\zb - z}\right)^4 \frac{z^4}{120}\partial^3_z z^2  \right] \nonumber\\&+ (z\leftrightarrow\zb),
\end{align}
and $h^{(k)}(z)$ are linear combinations of MPLs with rational function coefficients \cite{Bissi:2020woe}. We will provide the explicit expression of $h^{(4)}(z)$ in the discussion of three-loop leading log in \eqref{eq:h4}. 

The leading log parts \eqref{eq:llk} consist of MPLs, but in general they are not single-valued. To uplift the result to a single-valued function, we use the zigzag integrals $Z^{(k)}$ as suggested in \cite{Drummond:2012bg,Drummond:2022dxw}. The zigzag integrals are a series of integrals obeying
\begin{align}
    z\zb\partial_z\partial_\zb Z^{(k)}(z,\zb) =& Z^{(k-1)}(1-z,1-\zb),\\
    Z^{(1)}(z,\zb) =& W_2(z,\zb).
\end{align}
These integrals are weight $2k$ SVMPLs, and when $U\to 0$ they behave like 
\begin{equation}
    Z^{(k)}\left(z,\zb\right) \sim \log^k U ,\quad k\geq 2.
\end{equation}
The first few of them are \footnote{Notice that the SVMPLs $\mathcal{L}$ defined in \cite{Brown:2004ugm} are related to ours by $\mathcal{L}_{\vec{a}}=(-1)^{\#}G^{\rm sv}_{\vec{a}}$, where $\#$ stands for the number of non-zero parameters in ${\vec{a}}$. }
\begin{align}
    Z^{(1)}(z,\zb)&=-G^{\rm sv}_{2}+G^{\rm sv}_{10}  \; , \\
    Z^{(2)}(z,\zb)&=-G^{\rm sv}_{200}+G^{\rm sv}_{30} \; , \\
    Z^{(3)}(z,\zb)&=-G^{\rm sv}_{2210}+G^{\rm sv}_{2120}-2\zeta_{3}(-3G^{\rm sv}_{20}+2G^{\rm sv}_{21}) \; ,\\
    Z^{(4)}(z,\zb)&=-G^{\rm sv}_{2230}+G^{\rm sv}_{2320}-4\zeta_{3}(G^{\rm sv}_{23} - G^{\rm sv}_{220}) +20\zeta_{5} G^{\rm sv}_{20} \; ,
\end{align}
where all the parameters in $G^{\rm sv}_{\vec{a}}$ greater than $1$ are shorthand for
\begin{equation}
    n \equiv \underbrace{00\cdots0}_{n-1}1.
\end{equation}
It was shown in \cite{Drummond:2022dxw} that the leading log part of the correlator can be captured in a single-valued way using zigzag integrals and their derivatives 
\begin{equation}
    Z_m^{(k)}(z,\zb) = \underbrace{\cdots(1-z)\partial_z(-z\partial_z)(1-z)\partial_z(-z\partial_z)}_{m \text{ derivatives}}Z^{(k)}(z,\zb)\, .
\end{equation}
Our seed functions $W_{2}$ and $W_4$ also fall into the category of zigzag integrals. Actually, 
\begin{align}
    W_2(z,\zb) &= Z^{(1)}(z,\zb) \; , \\
    W_{4}(z,\zb) &= Z^{(2)}(1-1/z,1-1/\zb)\; .
\end{align}
We will soon show how to use the zigzag integrals to construct seed functions at higher loops. 

\subsection{The two-loop correlator}

We first discuss the correlator $\langle 2222 \rangle_\mathrm{GR}$ at the two-loop level. This correlator is bootstrapped by an ansatz inspired by the leading log structure \eqref{eq:llk}. The result can be written in a compact form through $\Delta^{(8)}$
\begin{equation}
    \mathcal{H}^{(3)}_{\mathrm{GR},2222} = \big[ \Delta^{(8)} \big]^2 \mathcal{L}^{(3)}_{\mathrm{GR},2222} + \frac{5}{4}\mathcal{H}^{(2)}_{\mathrm{GR},2222} - \frac{1}{16}\mathcal{H}^{(1)}_{\mathrm{GR},2222}\, .
\end{equation}
The function $\mathcal{L}^{(3)}_{\mathrm{GR},2222}$ is called the pre-correlator, which is a much simpler object comparing with the full correlator $\mathcal{H}^{(3)}_{\mathrm{GR},2222}$. The pre-correlator enjoys an accidental crossing symmetry
\begin{equation}\label{eq:L3crossing}
    \mathcal{L}^{(3)}(z,\zb) = \mathcal{L}^{(3)}(1-z,1-\zb) = \mathcal{L}^{(3)}\left(\frac{z}{z-1},\frac{\zb}{\zb-1}\right),
\end{equation}
and takes the following form in position space
\begin{equation}
    \mathcal{L}^{(3)} = \sum_{w=0}^6\sum_{s=\pm} \sum_i \frac{p_{w,s,i}(z,\zb)}{(z-\zb)^7} G_{w,s,i}(z,\zb)\, .
\end{equation}
Here $p_{w,s,i}(z,\zb)$ are some polynomials, and $G_{w,s,i}(z,\zb)$ are some SVMPLs, with $w$ and $s$ indicating the weight and parity respectively. The label $i$ is to distinguish independent SVMPLs with the same weight and parity. We will not present the explicit expression of $p_{w,s,i}(z,\zb)$ and $G_{w,s,i}(z,\zb)$ here, and refer to \cite{Huang:2021xws} for the full result.

Since $\mathcal{H}^{(2)}$ and $\mathcal{H}^{(1)}$ admit a differential representation, and $\Delta^{(8)}$ is already a differential operator, it is sufficient to focus on finding a differential representation for $\mathcal{L}^{(3)}$. This function $\mathcal{L}^{(3)}$ is still too complicated though, and we simply work on the highest weight part and the leading log part of it, and leave the full analysis for future work.

According to the crossing symmetry property \eqref{eq:L3crossing}, we define the Mellin amplitude of $\mathcal{L}^{(3)}$ to be \footnote{This integral seems to be ill-defined for $S=T=0$, where three sets of poles in the $\Gamma$ function collide together and pinch the integral contour. However, we can always subtract a constant $A_0$ from the pre-correlator $\mathcal{L}^{(3)}$ to eliminate the $U^0V^0$ term, without changing the full correlator $\mathcal{H}^{(3)}$. This makes the Mellin amplitude $\widetilde{\mathcal{M}}^{(3)}$ vanish at $S=T=0$, rendering the integral to be well-defined. }
\begin{align}
    \mathcal{L}^{(3)}(U,V) = \int \frac{\dd S\dd T}{(2\pi i)^2}U^S V^T \widetilde{\mathcal{M}}^{(3)}(S,T)\Gamma^2(-S)\Gamma^2(-T)\Gamma^2(-\tilde{U})\; ,
\end{align}
with $S+T+\tilde{U}=0$. By counting the power of $\log U$ and $\log V$ in $\mathcal{L}^{(3)}$, we expect the Mellin amplitude $\widetilde{\mathcal{M}}^{(3)}$ to have the following pole structure 
\begin{align}
    \widetilde{\mathcal{M}}^{(3)}(S,T)\ \sim&\ \frac{d_{m,n}}{(S-m)^2(T-n)} + \frac{e_{m,n}}{(S-m)(T-n)} + \frac{f_{m}}{(S-m)^2} + \frac{g_{m}}{(S-m)} \nonumber\\
    & + (\text{crossing}).
\end{align}
By ``+(crossing)'' we mean to add up terms under the permutations on $S,T,\tilde{U}$. In principle, these residues can be obtained from the series expansion of $\mathcal{L}^{(3)}$, through the contour deformation method described in \cref{sec:tree}. However, as we have mentioned above, it is too complicated to perform the full analysis on $\mathcal{L}^{(3)}$, and we only present parts of the residues in the following. 

One important aspect of this Mellin amplitude is that, unlike those amplitudes at the one-loop level, there will be an overall denominator $(S+T)^2$ in the standard form of it  
\begin{equation}
    \frac{1}{(S+T)^2}\widetilde{\mathcal{M}}^{(3)}(S,T),
\end{equation}
which requires that each term in $\widetilde{\mathcal{M}}^{(3)}(S,T)$ should have at least a double zero $S^2$, $T^2$ or $(S+T)^2$, as we will see in the following.

\subsubsection{Seed functions from the highest weight terms}

For a holographic correlator, the highest weight part is usually the simplest part. This is especially true for differential representation, since we can directly read off the seed functions needed from the position space result, as discussed in \cref{sec:diffdiscussion}. In $\mathcal{L}^{(3)}$, the highest weight is 6, and the corresponding results are given by two parity odd SVMPLs
\begin{align}
    \mathcal{L}^{(3)} \supset& \left(-\frac{P(U,V)}{960 (z-\zb)^7}\, W_{6,1}(z,\zb)  +(\text{crossing}) \right) + \frac{Q(U,V)}{13440 (z-\zb)^5}\, W_{6,2}(z,\zb)\, ,
\end{align}
where 
\begin{align}
    P(U,V) =&\ U^4 \left(6 + U^3 + U^2 (13 - 3 V) - 12 V + 7 V^2 - V^3 + U (22 - 20 V + 3 V^2)\right), \\
    Q(U,V) =&\ U^5 + U^4 (1 + V) + (-1 + V)^2 (1 + V)^3 - 2 U^3 (1 - 4 V + V^2) \nonumber\\
    & - 2 U^2 (1 + V + V^2 + V^3) + U (1 + 8 V - 2 V^2 + 8 V^3 + V^4)\; ,
\end{align}
and ``+(crossing)'' here means to plus five other terms related by the crossing symmetry
\begin{equation}
    z \to 1-z,\ \frac{1}{z},\ \frac{z}{z-1},\ \frac{1}{1-z},\ \frac{z-1}{z}, \text{ (so for $\zb$)}.
\end{equation}
The functions $W_{6,1}$ and $W_{6,2}$ are
\begin{align}
	W_{6,1}\left(z,\zb\right) =& -Z^{(3)}(1/z,1/\zb)\; , \\
	W_{6,2}\left(z,\zb\right) =&\ -G^{\rm sv}_{222}+G^{\rm sv}_{2210}+G^{\rm sv}_{213}-G^{\rm sv}_{2120}+G^{\rm sv}_{132}-G^{\rm sv}_{1310} -G^{\rm sv}_{123}+G^{\rm sv}_{1220} \nonumber\\
	&  -2 \zeta_3 (W_{3}(z,\zb) - W_3(1-z,1-\zb) - W_3(1/z,1/\zb) +G^{\rm sv}_{3}+3 G^{\rm sv}_{20}  \nonumber\\
	& -2G^{\rm sv}_{21}-G^{\rm sv}_{100}-G^{\rm sv}_{110}   ) + 15 \zeta_5 G^{\rm sv}_{1}\, .
\end{align}
The $W_{6,1}$ contributes to $\log^3 U$, the leading log part of $\mathcal{L}^{(3)}$. The $W_{6,2}$, instead, only contributes to the sub-leading log part $\log^2 U$. It is also a fully crossing symmetric function up to parity of permutations
\begin{equation}
    W_{6,2}(z,\zb) = -W_{6,2}(1-z,1-\zb) = -W_{6,2}(1/z,1/\zb).
\end{equation}

As with the other seed functions previously encountered, at this level we specify the following seed functions
\begin{align}
    \mathcal{W}_{6,1}(z,\bar{z})\equiv\frac{W_{6,1}(z,\zb)}{z-\zb} =& \int \frac{\dd S \dd T}{(2\pi i)^2} U^S V^T\, \widetilde{\Gamma}(S,T)\, \mathcal{M}_{6,1}(S,T), \label{eq:W61Mellin} \\
	\mathcal{W}_{6,2}(z,\bar{z})\equiv\frac{W_{6,2}(z,\zb)}{z-\zb} =& \int \frac{\dd S \dd T}{(2\pi i)^2} U^S V^T\, \widetilde{\Gamma}(S,T)\, \mathcal{M}_{6,2}(S,T). \label{eq:W62Mellin}
\end{align}
Unfortunately, the corresponding Mellin amplitudes $\mathcal{M}_{6,1}(S,T)$ and $\mathcal{M}_{6,2}(S,T)$ can no longer be written in a closed form in terms of harmonic sums anymore. Nevertheless, we can still express them as (divergent) infinite sums. Here we simply present their residues on simultaneous poles 
\begin{align}
    \mathcal{M}_{6,1}(S,T) \sim&\, \frac{A_{m,n}}{(S-m)^2(T-n)} + \frac{B_{m,n}}{(S-m)(T-n)}\; ,\\
    \mathcal{M}_{6,2}(S,T) \sim&\, \frac{C_{m,n}}{(S-m)(T-n)} + \frac{C_{m,n}}{(S-m)(\tilde{U}-n)} + \frac{C_{m,n}}{(T-m)(\tilde{U}-n)}\; ,
\end{align}
where $S+T+\tilde{U}=-1$, and
\begin{align}
    A_{m,n} =& S_1(m+n)-S_1(m),\\ \label{eq:Amn}
    B_{m,n} =& 2S_2(m) + 2S_1(m)S_1(m+n)-2S_{1,1}(m)-2S_{1,1}(m+n),\\
    C_{m,n} =& \big(S_1(m+n)-S_1(m)\big)\big(S_1(m+n)-S_1(n)\big).
\end{align}
These residues $A_{m,n}$, $B_{m,n}$, $C_{m,n}$ indeed show up in the Mellin amplitude of $\mathcal{L}^{(3)}$, with appropriate shifts, denominators and double zeros. 

By now we present the highest weight part of the $S^2T$ and $ST$ pole residues here
\begin{align}
    \widetilde{\mathcal{M}}^{(3)}\, \supset\ \frac{d_{m,n}}{(S-m)^2(T-n)} + \frac{e_{m,n}}{(S-m)(T-n)}
\end{align}
where
\begin{align}
    d_{m,n} =& -\frac{7 (m-3)^2 (m-2)^2 (m-1)^2 m^2}{38400 (n+3)}A_{m-4,n+3} \nonumber\\
    &+\frac{(m-3) (m-2)^2 (m-1)^2 m^2 (7 m-17)}{19200 (n+2)}A_{m-3,n+2} \nonumber\\
    & -\frac{(m-3) (m-2) (m-1)^2 m^2 \left(21 m^2-81 m+80\right)  }{115200 (n+1)}A_{m-2,n+1} \nonumber\\
    & + (\text{lower weights}),\\[5mm]
    e_{m,n} =& -\frac{7 (m-3)^2 (m-2)^2 (m-1)^2 m^2}{38400 (n+3)}B_{m-4,n+3} \nonumber\\
    &+\frac{(m-3) (m-2)^2 (m-1)^2 m^2 (7 m-17)}{19200 (n+2)}B_{m-3,n+2} \nonumber\\
    & -\frac{(m-3) (m-2) (m-1)^2 m^2 \left(21 m^2-81 m+80\right)  }{115200 (n+1)}B_{m-2,n+1} \nonumber\\
    & + (m\leftrightarrow n) \nonumber\\[2mm]
    & + \frac{m^2 \left(6 m^2-8 m+3\right)}{40320} C_{m-1,n} + \frac{n^2 \left(6 n^2-8 n+3\right)}{40320} C_{m,n-1} \nonumber\\
    & + \frac{(m+n)^2 \left(6 (m+n)^2+8 (m+n)+3\right)}{40320} C_{m,n}  \nonumber\\
    & + (\text{lower weights}).
\end{align}
This form of residues precisely match the requirement of differential representation. From the residues above we can conclude that, the highest weight part of the Mellin amplitude $\widetilde{\mathcal{M}}^{(3)}$ are given by
\begin{align}
    \widetilde{\mathcal{M}}^{(3)} \supset &-\frac{7 (S-3)^2 (S-2)^2 (S-1)^2 S^2}{38400 (T+3)}\mathcal{M}_{6,1}(S-4,T+3) \nonumber\\
    &+\frac{(S-3) (S-2)^2 (S-1)^2 S^2 (7 S-17)}{19200 (T+2)}\mathcal{M}_{6,1}(S-3,T+2) \nonumber\\
    & -\frac{(S-3) (S-2) (S-1)^2 S^2 \left(21 S^2-81 S+80\right)  }{115200 (T+1)}\mathcal{M}_{6,1}(S-2,T+1) \nonumber\\
    & + \frac{1}{2}\frac{S^2 \left(6 S^2-8 S+3\right)}{40320} \mathcal{M}_{6,2}(S-1,T) \nonumber\\[2mm]
    & + (\text{crossing})\; ,
\end{align}
and the differential operators can be written down accordingly
\begin{align}\label{eq:L3hw}
    \mathcal{L}^{(3)} \supset &\ \mathcal{D}_{6,1} \mathcal{M}_{6,1}(z,\zb) + \frac{1}{80640}(3-8\DU+6\DU^2)U \mathcal{M}_{6,2}(z,\zb) \nonumber\\
    & + (\text{crossing})\; ,
\end{align}
where
\begin{align}
    \mathcal{D}_{6,1} &= -\frac{7}{38400}(1+\DV)^2(2+\DV)^2(3+\DV)U^4V^{-3} \nonumber\\
    &\quad  + \frac{1}{19200}(-3+\DU)(-17+7\DU)(1+\DV)^2(2+\DV)U^3V^{-2} \nonumber\\
    &\quad  - \frac{1}{115200}(-3+\DU)(-2+\DU)(80-81\DU+21\DU^2)(1+\DV)U^2V^{-1}.
\end{align}
By comparing with the full expression of $\mathcal{L}^{(3)}$, one can check that \eqref{eq:L3hw} accounts for all weight 6 SVMPLs correctly. In addition, by subtracting \eqref{eq:L3hw} in $\mathcal{L}^{(3)}$, we also find the parity even weight 5 SVMPLs disappear. This is exactly what we expect for differential representation. 

\subsubsection{The leading log part in differential representation}

We now turn to the leading log part of $\mathcal{L}^{(3)}$. Following \cite{Drummond:2022dxw}, we will use the zigzag integral $Z^{(3)}$ (or equivalently $W_{6,1}$) and its derivatives to form our basis of seed functions. However, one difference is that we only admit the SVMPLs with odd parity. These functions are
\begin{align}
    W_{5,1}(z,\zb) =& -z\partial_z W_{6,1}(z,\zb) - (z\leftrightarrow\zb), \\
    W_{4,1}(z,\zb) =& -z(1-z)\partial_z z\partial_z W_{6,1}(z,\zb) - (z\leftrightarrow\zb) - \zeta_2 W_2(z,\zb).
\end{align}
The $- \zeta_2 W_2$ in the definition of $W_{4,1}$ is to ensure the Mellin amplitude of $\mathcal{W}_{4,1}$ to be $S_2(-S-1)$. The Mellin amplitudes are related to the seed functions by the standard Mellin transformation  
\begin{align}
    \mathcal{W}_{5,1}(z,\bar{z})\equiv\frac{W_{5,1}(z,\zb)}{z-\zb} =& \int \frac{\dd S \dd T}{(2\pi i)^2} U^S V^T\, \widetilde{\Gamma}(S,T)\, \mathcal{M}_{5,1}(S,T), \label{eq:W51Mellin} \\
    \mathcal{W}_{4,1}(z,\bar{z})\equiv\frac{W_{4,1}(z,\zb)}{z-\zb} =& \int \frac{\dd S \dd T}{(2\pi i)^2} U^S V^T\, \widetilde{\Gamma}(S,T)\, S_2(-S-1),
\end{align}
where the residues of $\mathcal{M}_{5,1}(S,T)$ on simultaneous poles are given by
\begin{equation}
    \mathcal{M}_{5,1}(S,T) \sim \frac{1}{(S-m)^2(T-n)} + \frac{-2A_{m,n}}{(S-m)(T-n)},
\end{equation}
and $A_{m,n}$ is the same as defined in \eqref{eq:Amn}. To match the leading log part of $\mathcal{L}^{(3)}$, we should use $\mathcal{M}_{6,1}$, $\mathcal{M}_{5,1}$, $\mathcal{M}_{4,1}$ and their crossing transformations to match all double pole terms in the Mellin amplitude $\widetilde{\mathcal{M}}^{(3)}$. This can be done by first considering the residues on $S^2T$ poles \footnote{A similar object, the $S^2T$ pole residues in the Mellin amplitude of $\mathcal{H}^{(3)}$, has been computed in \cite{Bissi:2020woe}. One can check that the residues there can be collected in a clearer form as in \eqref{eq:dmn} and satisfy the double zero property. }
\begin{align}\label{eq:dmn}
    d_{m,n} =& -\frac{7 (m-3)^2 (m-2)^2 (m-1)^2 m^2}{38400 (n+3)}A_{m-4,n+3} \nonumber\\
    &+\frac{(m-3) (m-2)^2 (m-1)^2 m^2 (7 m-17)}{19200 (n+2)}A_{m-3,n+2} \nonumber\\
    & -\frac{(m-3) (m-2) (m-1)^2 m^2 \left(21 m^2-81 m+80\right)  }{115200 (n+1)}A_{m-2,n+1} \nonumber\\[2mm]
    & + \frac{(m-3)^2 (m-2)^2 (m-1)^2 m^2}{3456000 (m+n-3)}-\frac{(m-3) (m-2)^2 (m-1)^2 m^2 (2 m+13)}{3456000 (m+n-2)} \nonumber\\
    & +\frac{(m-3) (m-2) (m-1)^2 m^2 \left(m^2+14m+105\right)}{3456000 (m+n-1)}.
\end{align}
From the residue $d_{m,n}$ we can construct the differential representation of $\mathcal{M}_{6,1}$ and $\mathcal{M}_{5,1}$ in $S^2T$ channel, and use crossing transformation to cover all other $S^2\tilde{U}$,$T^2S$, $T^2\tilde{U}$, $\tilde{U}^2S$, $\tilde{U}^2T$ poles. We have
\begin{align}\label{eq:W61W51}
    \mathcal{L}^{(3)} \supset &\ \mathcal{D}_{6,1} \mathcal{M}_{6,1}(z,\zb) + \mathcal{D}_{5,1} \mathcal{M}_{5,1}(z,\zb) +\ (\text{crossing})\; ,
\end{align}
where
\begin{align}
    \mathcal{D}_{5,1} =&\ \frac{1}{3456000}(-1+\DU+\DV)^2(-2+\DU+\DV)^2(-3+\DU+\DV)U^4  \nonumber\\
    &  - \frac{1}{3456000}(-3+\DU)(13+2\DU)(-1+\DU+\DV)^2(-2+\DU+\DV)U^3 \nonumber\\
    &  + \frac{1}{3456000}(-3+\DU)(-2+\DU)(105+14\DU+21\DU^2)(-1+\DU+\DV)U^2.
\end{align}
Then we subtract \eqref{eq:W61W51} from $\mathcal{L}^{(3)}$, and consider the Mellin amplitude of the remaining function. By construction the only double poles in it are $S^2$, $T^2$ and $\tilde{U}^2$ poles. The Laurent series coefficient for $(S-m)^{-2}(T-n)^0$ in this Mellin amplitude can be worked out as
\begin{align}
    & -\frac{7 (m-3) (m-2)^2 (m-1)^2 m^2}{38400} \left(\frac{1}{m+n-2}-\frac{1}{n+2}\right) \nonumber\\
    & + \frac{(m-3) (m-2) (m-1)^2 m^2 (14 m-33) }{76800}\left(\frac{1}{m+n-1}  -\frac{1}{n+1}\right) .
\end{align}
Comparing with the Mellin amplitude of $\mathcal{W}_{4,1}(z,\bar{z})$, we find that these coefficients come from
\begin{align}
    &\quad \frac{7 (S-3) (S-2)^2 (S-1)^2 S^2}{38400} \left(\frac{1}{S+T-2}-\frac{1}{T+2}\right)S_2(-S-1) \nonumber\\
    & -\frac{(S-3) (S-2) (S-1)^2 S^2 (14 S-33) }{76800}\left(\frac{1}{S+T-1}  -\frac{1}{T+1}\right)S_2(-S-1).
\end{align}
Therefore we can write down our Mellin amplitude for the leading log part and the corresponding differential representation as
\begin{align}
    \widetilde{\mathcal{M}}^{(3)} \supset& -\frac{7 (S-3)^2 (S-2)^2 (S-1)^2 S^2}{38400 (T+3)}\mathcal{M}_{6,1}(S-4,T+3) \nonumber\\
    &+\frac{(S-3) (S-2)^2 (S-1)^2 S^2 (7 S-17)}{19200 (T+2)}\mathcal{M}_{6,1}(S-3,T+2) \nonumber\\
    & -\frac{(S-3) (S-2) (S-1)^2 S^2 \left(21 S^2-81 S+80\right)  }{115200 (T+1)}\mathcal{M}_{6,1}(S-2,T+1) \nonumber\\[2mm]
    & + \frac{(S-3)^2 (S-2)^2 (S-1)^2 S^2}{3456000 (S+T-3)}\mathcal{M}_{5,1}(S-4,T) \nonumber\\
    & -\frac{(S-3) (S-2)^2 (S-1)^2 S^2 (2 S+13)}{3456000 (S+T-2)}\mathcal{M}_{5,1}(S-3,T) \nonumber\\
    & +\frac{(S-3) (S-2) (S-1)^2 S^2 \left(S^2+14S+105\right)}{3456000 (S+T-1)}\mathcal{M}_{5,1}(S-2,T) \nonumber\\[2mm]
    & -\frac{7 (S-3) (S-2)^2 (S-1)^2 S^2}{38400(T+2)}S_2(-S+1) \nonumber\\
    & + \frac{(S-3) (S-2) (S-1)^2 S^2 (14 S-33) }{76800(T+1)}S_2(-S) \nonumber\\
    & + (\text{crossing})\; ,
\end{align}
and 
\begin{align}\label{eq:L3LL}
    \mathcal{L}^{(3)} \supset &\ \mathcal{D}_{6,1} \mathcal{M}_{6,1}(z,\zb) + \mathcal{D}_{5,1} \mathcal{M}_{5,1}(z,\zb) + \mathcal{D}_{4,1} \mathcal{W}_{4,1}(z,\zb) \nonumber\\
    & + (\text{crossing})\; ,
\end{align}
where
\begin{align}
    \mathcal{D}_{4,1} =& - \frac{7}{38400}(-3+\DU)(1+\DV)^2(2+\DV)U^3V^{-2} \nonumber\\
    & + \frac{1}{76800}(-3+\DU)(-2+\DU)(-33+14\DU)(1+\DV)U^2V^{-1}.
\end{align}
The leading log part of the full correlator $\mathcal{H}^{(3)}$ can be obtained by acting $\big[ \Delta^{(8)} \big]^2$ 
\begin{align}\label{eq:H3LL}
    \mathcal{H}^{(3)} \supset &\ \big[ \Delta^{(8)} \big]^2 \times \Big( \mathcal{D}_{6,1} \mathcal{M}_{6,1}(z,\zb) + \mathcal{D}_{5,1} \mathcal{M}_{5,1}(z,\zb) + \mathcal{D}_{4,1} \mathcal{W}_{4,1}(z,\zb) \Big) \nonumber\\
    & + (\text{crossing})\; .
\end{align}

\subsubsection{The leading log correlator}

Comparing with the leading log part $\mathcal{G}^{(3)}$ constructed in \cite{Bissi:2020wtv}, our leading log expression \eqref{eq:L3LL} contains a lot more information. The $\mathcal{G}^{(3)}$ there can be viewed as the result of acting the differential operators purely on $1/(z-\zb)$ in the seed functions. In fact, the leading log expression \eqref{eq:L3LL} itself can be regarded as a correlator, because it satisfies all the properties of a conformal correlator (crossing symmetry, single-valuedness, and the finiteness at $z=\zb$). Since it also captures all the leading log information, we call \eqref{eq:L3LL} and \eqref{eq:H3LL} the \emph{leading log (pre-)correlator} of $\langle 2222 \rangle_\mathrm{GR}$ at the two-loop level. It would be interesting to explore if the leading log correlator describes any subsector of operators in the original conformal field theory. 

The concept of leading log correlator may be particularly useful at one loop, where there are only three seed functions $\mathcal{W}_{2,3,4}$, and two of them contribute to the leading log part. By construction, the difference between the leading log correlator and the correlator itself is just given by some contact diagrams. 

Recently, Heslop, Lipstein and Santagata provide a similar construction for ${\rm AdS_2} \times {\rm S}^2$ in \cite{Heslop:2023gzr} from a different starting point. They find that, by treating ${\rm AdS}$ and S on an equal footing, it is possible to derive the 4d hidden conformal symmetry in ${\rm AdS_2 }\times {\rm S}^2$ at the one-loop level. They provide a 4d uplift for all the correlators of chiral primaries
\begin{align}\label{eq:ads2}
	& {\bf G}^{(2)} = -\frac{1}{2}\Big( 
	  \mathcal{C}_{12} \frac{\mathfrak{f}_s(\mathbf{z},\mathbf{\bar{z}}) }{\mathbf{x}_{13}^2 \mathbf{x}_{24}^2}  + \mathcal{C}_{23} \frac{\mathfrak{f}_t(\mathbf{z},\mathbf{\bar{z}})}{\mathbf{x}_{13}^2 \mathbf{x}_{24}^2} +  \mathcal{C}_{13} \frac{\mathfrak{f}_u(\mathbf{z},\mathbf{\bar{z}}) }{\mathbf{x}_{13}^2 \mathbf{x}_{24}^2} \Big)\,,
\end{align}
where the function $\mathfrak{f}_s(\mathbf{z},\mathbf{\bar{z}})$, quite remarkably, is exactly equal to the seed function $\mathcal{W}_{3}(\mathbf{z},\mathbf{\bar{z}})$ in our paper! The operator $\mathcal{C}_{12}$ here is an ${\rm AdS_2 }\times {\rm S}^2$ analogy of $\Delta^{(8)}$ in ${\rm AdS_5 }\times {\rm S}^5$, and the $\mathcal{C}_{12} \mathbf{x}_{13}^{-2} \mathbf{x}_{24}^{-2}\mathfrak{f}_s(\mathbf{z},\mathbf{\bar{z}})$ there is fixed by matching with the leading log of the one-loop correlator. Therefore, for the ${\rm AdS_2 \times {\rm S}^2}$ cases, the leading log correlator constructed in a similar way as above will provide the correct result directly, while keeping the 4d hidden conformal symmetry manifest. We believe that, our construction here gives the higher dimensional generalization of \eqref{eq:ads2} in ${\rm AdS_5} \times {\rm S}^5$ and ${\rm AdS_5} \times {\rm S}^3$. It would be interesting to investigate this topic in details, and we leave it for future.

\subsection{Three-loop leading log}

In this subsection, we briefly show how to construct the leading log correlator at the three-loop level using differential representation. Our starting point is the zigzag integral $Z^{(4)}$, or $W_{8}$ in our convention
\begin{equation}
    W_{8}(z,\zb) = Z^{(4)}(1-1/z,1-1/\zb)
\end{equation}
and the derivatives of it
\begin{align}
    W_{7}(z,\zb) =& -z(1-z)\partial_z W_{8}(z,\zb) - (z\leftrightarrow\zb) ,\\
    W_{6}(z,\zb) =&\ z\partial_z z(1-z)\partial_z W_{8}(z,\zb) - (z\leftrightarrow\zb) ,\\
    W_{5}(z,\zb) =& -z(1-z)\partial_z z\partial_z z(1-z)\partial_z W_{8}(z,\zb) - (z\leftrightarrow\zb) .
\end{align}
The seed functions and their Mellin amplitudes are defined as before
\begin{equation}
    \mathcal{W}_{i}(z,\bar{z})\equiv\frac{W_{i}(z,\zb)}{z-\zb} = \int \frac{\dd S \dd T}{(2\pi i)^2} U^S V^T\, \widetilde{\Gamma}(S,T)\, \mathcal{M}_{i}(S,T),\quad i=5,6,7,8. \label{eq:WiMellin}
\end{equation}
We record the $S^3T$ pole residues of $\mathcal{M}_{8/7/6}(S,T)$, and the $S^3$ pole residues of $\mathcal{M}_{5}(S,T)$ below 
\begin{align}
    \mathcal{M}_{8}(S,T) \sim&\ \frac{D_{m,n}}{(S-m)^3(T-n)},\\
    \mathcal{M}_{7}(S,T) \sim&\ \frac{A_{m,n}}{(S-m)^3(T-n)},\\
    \mathcal{M}_{6}(S,T) \sim&\ \frac{1}{(S-m)^3(T-n)},\\
    \mathcal{M}_{5}(S,T) \sim&\ \frac{1}{(S-m)^3},
\end{align}
where
\begin{align}
    D_{m,n} =&\ \widetilde{S}(m,n) - S_1(m)S_1(m+n)+S_1(n)S_1(m+n)+S_2(m+n)  \nonumber\\
    & +S_{1,1}(m)-S_{1,1}(m+n),
\end{align}
and $\widetilde{S}(m,n)$ here is a two-variable generalization of harmonic sums
\begin{equation}
    \widetilde{S}(m,n) = \sum_{i=1}^{m-1}\frac{S_1(i)}{m+n-i}.
\end{equation}

The leading log part at three loops is given by 
\begin{equation}
    \mathcal{H}^{(4)}\Big|_{\log^4 U} = \big[ \Delta^{(8)} \big]^3 \mathcal{D}_{(3)} h^{(4)}(z),
\end{equation}
with
\begin{align} \label{eq:h4}
    h^{(4)}(z) =&\ \frac{3231098431 z^4-9277069595 z^3+8883295360 z^2-2955774240 z+170794800}{7166361600000 z^4} \nonumber\\
    & + \frac{(z-1) \left(1604798 z^3-3262153 z^2+1770392 z-143277\right)}{9953280000 z^5}G_{1,z} \nonumber\\
    & -\frac{(z-1) \left(802399 z^4-2276176 z^3+2137824 z^2-683176 z+34249\right)}{4976640000 z^5}\big(G_{01,z}-G_{11,z}\big) \nonumber\\
    & +\frac{\left(8130 z^4-22185 z^3+19595 z^2-5470 z+182\right)}{82944000 z^5}\big(G_{001,z}-G_{011,z}\big) \nonumber\\
    & -\frac{(z-1)^3 \left(271 z^2-62 z+1\right)}{2764800 z^5} \big(G_{0001,z}-G_{0011,z}-G_{1001,z}+G_{1011,z}\big) \nonumber\\[1mm]
    & - \left( z\leftrightarrow \frac{z}{z-1} \right).
\end{align}
Although $\mathcal{D}_{(3)} h^{(4)}(z)$ is not single-valued, we can still consider a pre-correlator-like object $\mathcal{L}^{(4)}$, which satisfy that
\begin{equation}
    \mathcal{L}^{(4)}\Big|_{\log^4 U} = \mathcal{D}_{(3)} h^{(4)}(z)\, ,
\end{equation}
and define the corresponding Mellin amplitude to be 
\begin{equation}
    \mathcal{L}^{(4)} =  \int \frac{\dd S\dd T}{(2\pi i)^2}U^{S} V^T \widetilde{\mathcal{M}}^{(4)}(S,T)\Gamma^2(-S)\Gamma^2(-T)\Gamma^2(S+T)\; .
\end{equation}
Here, only the leading log part of $\mathcal{L}^{(4)}$ and the $S^3$ pole part of $\widetilde{\mathcal{M}}^{(4)}(S,T)$ are relevant to our discussion. Comparing with the $\log^4U\log^2V$ term in $\mathcal{L}^{(4)}$, we obtain the $S^3T$ pole residues
\begin{align}
    \widetilde{\mathcal{M}}^{(4)}(S,T) \sim\ \frac{h_{m,n}}{(S-m)^3(T-n)},
\end{align}
where
\begin{equation}
    h_{m,n}= h^D_{m,n} + h^A_{m,n} + h^1_{m,n}
\end{equation}
with
\begin{align}\label{eq:hmn}
   h^D_{m,n}=&\ \frac{271 (m-3)^2 (m-2)^2 (m-1)^2 m^2 }{27648000 (m+n-3)}D_{m-4,n} \nonumber\\
   &-\frac{(m-3) (m-2)^2 (m-1)^2 m^2 (271 m-751)  }{13824000 (m+n-2)}D_{m-3,n} \nonumber\\
   &+\frac{(m-3) (m-2) (m-1)^2 m^2 \left(271 m^2-1231 m+1380\right)  }{27648000 (m+n-1)}D_{m-2,n} \\[1mm]
   h^A_{m,n}=& -\frac{7 (m-3)^2 (m-2)^2 (m-1)^2 m^2 }{23040000 (n+3)}A_{m-4,n+3} \nonumber\\
   &+\frac{7 (m-3) (m-2)^2 (m-1)^2 m^2 (4 m+1)  }{46080000 (n+2)}A_{m-3,n+2} \nonumber\\
   &-\frac{7 (m-3) (m-2) (m-1)^2 m^2 \left(18 m^2+27 m+215\right) }{414720000 (n+1)}A_{m-2,n+1} \\[1mm]
   h^1_{m,n}=&\ \frac{802399 (m-3)^2 (m-2)^2 (m-1)^2 m^2}{49766400000 (m+n-3)} \nonumber\\
   &-\frac{(m-3) (m-2)^2 (m-1)^2 m^2 (802399 m-1979719) }{24883200000 (m+n-2)} \nonumber\\
   &+\frac{(m-3) (m-2) (m-1)^2 m^2 \left(802399 m^2-3157039 m+2602020\right)}{49766400000 (m+n-1)}.
\end{align}
Following a similar procedure as at two loops, we can write down the differential representation of \eqref{eq:hmn} and the corresponding $S^3\tilde{U}$ pole contribution, and subtract them from $\mathcal{L}^{(4)}$. The Mellin amplitude of the remaining function has $S^3$ poles, and the Laurent series coefficient around $(S-m)^{-3}(T-n)$ is  
\begin{align}
    & -\frac{7 (m-3) (m-2)^2 (m-1)^2 m^2}{23040000} \left(\frac{1}{m+n-2}-\frac{1}{n+2}\right) \nonumber\\
    & + \frac{7 (m-3) (m-2) (m-1)^2 m^2 (m+3)}{23040000}\left(\frac{1}{m+n-1}  -\frac{1}{n+1}\right) .
\end{align}
Therefore, we can translate the above residues into differential operators, and write down the leading log correlator for $\mathcal{H}^{(4)}$ 
\begin{align}
    \mathcal{H}^{(4)} \supset& \big[ \Delta^{(8)} \big]^3 \times \Big(\mathcal{D}_8 \mathcal{W}_8(z,\zb)+\mathcal{D}_7 \mathcal{W}_7(z,\zb)+\mathcal{D}_6 \mathcal{W}_6(z,\zb)+\mathcal{D}_5 \mathcal{W}_5(z,\zb) \Big) \nonumber \\
    & + (\text{crossing})\; ,
\end{align}
where
\begin{align}
    \mathcal{D}_8 =&\ \frac{271}{27648000}\Delta_{-3,-1}U^4 - \frac{1}{13824000}(-3+\DU)(-751+271\DU)\Delta_{-2,-1}U^3  \nonumber\\
    & + \frac{1}{27648000}(-3+\DU)(-2+\DU)(1380-1231\DU+271\DU^2)\Delta_{-1,-1}U^2, \nonumber\\[2mm]
    \mathcal{D}_7 =&-\frac{7}{23040000}(1+\DV)^2(2+\DV)^2(3+\DV)U^4V^{-3} \nonumber\\
    & + \frac{7}{46080000}(-3+\DU)(1+4\DU)(1+\DV)^2(2+\DV)U^3V^{-2} \nonumber\\
    & - \frac{7}{414720000}(-3+\DU)(-2+\DU)(215+27\DU+18\DU^2)(1+\DV)U^2V^{-1}, \nonumber\\[2mm]
    \mathcal{D}_6 =&\ \frac{802399}{49766400000}\Delta_{-3,-1}U^4 - \frac{1}{24883200000}(-3+\DU)(-1979719+802399\DU)\Delta_{-2,-1}U^3 \nonumber\\
    & + \frac{1}{49766400000}(-3+\DU)(-2+\DU)(2602020-3157039\DU+802399\DU^2)\Delta_{-1,-1}U^2, \nonumber\\[2mm]
    \mathcal{D}_5 =&\ \frac{7}{23040000}(-3+\DU)(1+\DV)^2(2+\DV)U^3V^{-2} \nonumber\\
    & - \frac{7}{23040000}(-3+\DU)(-2+\DU)(3+\DU)(1+\DV)U^2V^{-1}.
\end{align}

\section{Discussion}\label{sec:discussion}

In this paper, we present a differential representation for holographic four-point correlators. In this approach, the correlators can be written as the result of acting differential operators on some seed functions with simple Mellin amplitudes. Using this method, one can directly go from the Mellin amplitude of a correlator to the position space expression and vice versa. We illustrate this idea by several examples in ${\rm AdS}_5\times {\rm S}^5$ and ${\rm AdS}_5\times {\rm S}^3$, including the supergravition correlator $\langle 2222 \rangle_\mathrm{GR}$ from the tree level to three-loop level,\footnote{At the one-loop level we also consider the stringy correction. At the two- and three-loop level we only present the partial results. } and the supergluon correlators $\langle 2222 \rangle_\mathrm{YM}$ and $\langle 3333 \rangle_\mathrm{YM}$ at the one-loop level. Through the differential representation, we discover an important feature, the double zero property, for Mellin amplitudes of holographic correlators, and verify this property in all the examples mentioned above. 

We stress that, although we only focus on four-point functions in ${\rm AdS}_5\times {\rm S}^5$ and ${\rm AdS}_5\times {\rm S}^3$, potentially this method can be generalized to higher-point functions and many other backgrounds. For example, it is already shown in \cite{Alday:2022lkk,Goncalves:2023oyx} that the tree-level five-point correlators $\langle 22222 \rangle_\mathrm{YM}$ in ${\rm AdS}_5\times {\rm S}^3$ and $\langle 222pp \rangle_\mathrm{GR}$ in ${\rm AdS}_5\times {\rm S}^5$ can be decomposed on the seed function $\bar{D}_{11112}$. As for other backgrounds, the tree-level four-point functions in ${\rm AdS}_{d+1}\times  {\rm S}^k$ with $d$ even can be written as sum over finitely many contact diagrams, and thus all of them can also be reduced to differentiations on $\bar{D}_{1111}$ in ${\rm AdS}_{d+1}$ \cite{Alday:2020dtb,Alday:2020lbp,Alday:2021odx}. For those backgrounds with $d$ odd, it seems like we need to include an additional seed function for the exchange Witten diagrams. It would be interesting to explore this formalism further for more complicated holographic correlators, like loop-level correlators in ${\rm AdS}_7\times {\rm S}^4$ and ${\rm AdS}_4\times {\rm S}^7$ \cite{Alday:2020tgi,Alday:2022rly}, or correlation functions higher than five-points in ${\rm AdS}_5\times {\rm S}^3$ \cite{Alday:2023kfm,Cao:2023cwa}. 

The method we present in this paper is in a empirical manner. We start from some known correlators, try to rewrite them in the differential representation, but we do not have an explanation on why they should take such forms. The Witten diagram descriptions of $\mathcal{W}_2$ and $\mathcal{W}_3$ suggest that there may be Witten diagram origins for all the seed functions. Moreover, there may exist a procedure that allows us to reduce any Witten diagrams on a specific seed function basis, just like the Feynman integral reduction in flat space scattering amplitudes. Despite the lacking of diagrammatic interpretation, the differential representation is proved to be very useful for one-loop correlators in ${\rm AdS}_5\times {\rm S}^5$ and ${\rm AdS}_5\times {\rm S}^3$. In a soon-to-appear work \cite{Huang:2024xxx} we will use this tool to derive all one-loop correlators $\langle p_1p_2p_3p_4 \rangle$ with extremality 2 for supergravitons and supergluons, with very limited demand on the physical data input.

We summarize several possible directions to explore the differential representation in the rest of this section:

\begin{itemize}
    \item It would be interesting to consider more AdS backgrounds such as ${\rm AdS}_2\times {\rm S}^2$ \cite{Abl:2021mxo,Heslop:2023gzr,Rigatos:2024yjo}, ${\rm AdS}_3\times {\rm S}^3$ \cite{Giusto:2018ovt,Rastelli:2019gtj,Giusto:2019pxc,Giusto:2020neo}, ${\rm AdS}_7\times {\rm S}^4$ \cite{Alday:2020tgi,Alday:2020dtb,Alday:2020lbp} and gluon scattering \cite{Alday:2021odx}. For theories with non-truncated spectrum such as ${\rm AdS}_4\times {\rm S}^7$ \cite{Alday:2020dtb,Alday:2021ymb,Alday:2022rly}, we need new seed functions to capture their special characteristics.
    \item Some recent works have made many progress on high-point tree-level amplitudes \cite{Goncalves:2019znr,Alday:2022lkk,Goncalves:2023oyx,Li:2023azu,Alday:2023kfm,Cao:2023cwa}, which motivates us to explore whether there is a similar correspondence between Mellin amplitudes and position space results.
    \item It is also interesting to explore other coupling regions, such as small 't Hooft coupling in the planar limit of $\mathcal{N}=4$ SYM, which is dual to the strong coupled type IIB theory in AdS. Some research \cite{Bhat:2023lev} has already been done on the perturbative position space results and their corresponding Mellin amplitudes.
    \item As mentioned before, the seed functions seem to have close connection with specific Witten diagrams. It is both interesting and crucial to gain a concrete understanding on this possible relation. This may provide more physical insights for the improvement of bootstrap computations.
\end{itemize}

\acknowledgments
	
ZH, BW and EYY are supported by National Natural Science Foundation of China under Grant No.~12175197 and Grand No.~12347103. EYY is also supported by National Natural Science Foundation of China under Grant No.~11935013, and by the Fundamental Research Funds for the Chinese Central Universities under Grant No.~226-2022-00216.

\appendix

\section{Proof of the identity \eqref{eq:HSidentity}}\label{appx:identity}

Here we will prove the identity \eqref{eq:HSidentity} used in the resummation of Mellin amplitudes in \cref{sec:gluon2222}. The identity reads
\begin{equation}\label{eq:hsid}
	S_1(z)S_1(-z-1)-S_{1,1}(z)-S_{1,1}(-z-1)-2\zeta_2 = 0.
\end{equation}
The simplest way to prove it is to write the harmonic sums in ploygamma functions 
\begin{align}
    S_1(z) =& \psi(z+1) + \gamma,\\
    S_{1,1}(z) =& \frac{1}{2}S_1^2(z) + \frac{1}{2}S_2(z) = \frac{1}{2}[\psi(z+1) + \gamma]^2 + \frac{1}{2}[-\psi'(z+1) + \zeta_2].
\end{align}
and use the reflection formula of polygammas 
\begin{align}
    \psi(-z) - \psi(z+1) =&\, \pi \cot \pi z,\\
    \psi'(-z) + \psi'(z+1) =&\, \pi^2 \csc^2 \pi z.
\end{align}
Substituting all these identities into \eqref{eq:hsid}, the left-hand-side turns to be
\begin{equation}
	\frac{\pi^2}{2}(\csc^2 \pi z - \cot^2 \pi z - 1),
\end{equation}
which is identically 0 and we prove \eqref{eq:hsid}.

\section{Resummation of $\widetilde{\mathcal{M}}^{(2)}_{\mathrm{YM},2222}(S,T)$}\label{appx:resum}

In this appendix, we perform a more rigorous computation of the double infinite sum
\begin{equation}
	\sum_{m,n=0}^{\infty} \frac{a_{m,n}}{(S-m)(T-n)},
\end{equation}
with 
\begin{equation}
	a_{m,n} = \frac{3 m^2 n+2 m^2+3 m n^2+8 m n+3 m+2 n^2+3 n}{3(m+n) (m+n+1) (m+n+2)}.
\end{equation}
We will first carry out the sum over $m$, which is convergent, and then the sum over $n$, which will diverge and can be regularized by taking derivatives on $T$. 

For the sum over $m$, we rewrite $a_{m,n}$ by partial fractions on $m$,
\begin{equation}
	a_{m,n} = -\frac{2 n^2}{3 (m+n)}+\frac{n^2+n+1}{3 (m+n+1)}+\frac{(n+1)^2}{3 (m+n+2)},
\end{equation}
and the sum on each terms can be worked out by using
\begin{equation}
    \sum_{m=0}^\infty \frac{1}{(m+a)(S-m)} = \frac{1}{S+a}\big[\psi(-S)-\psi(a)\big].
\end{equation}
The result becomes 
\begin{align}\label{eq:partialsum}
    \sum_{n=0}^\infty \frac{1}{T-n}&\left[  -\frac{2 n^2}{3 (S+n)}\big[\psi(-S)-\psi(n)\big]+\frac{n^2+n+1}{3 (S+n+1)}\big[\psi(-S)-\psi(n+1)\big]\right. \nonumber \\
    &\left. \quad +\frac{(n+1)^2}{3 (S+n+2)}\big[\psi(-S)-\psi(n+2)\big] \right].
\end{align}
The summand seems to be $\order{\log(n)}$ when $n\to\infty$, but there are cancellations between different terms and the summand actually goes like $\order{n^{-1}}$. It is still a divergent sum, but we can isolate the term of $\order{n^{-1}}$ by taking partial fractions on $n$ in the parenthesis in \eqref{eq:partialsum}, which gives
\begin{align}
    \sum_{n=0}^\infty \frac{1}{T-n}&\left[  \left(-\frac{2 S^2}{3 (S+n)}+\frac{S^2+S+1}{3 (S+n+1)}+\frac{(S+1)^2}{3 (S+n+2)} \right)\psi(-S) \right. \nonumber \\
    & \quad -\left(-\frac{2 S^2}{3 (S+n)}+\frac{S^2+S+1}{3 (S+n+1)}+\frac{(S+1)^2}{3 (S+n+2)} \right)\psi(n+1) \nonumber\\
    &\left. \quad + \frac{2 S}{3 (S+n)}+\frac{S+1}{3 (S+n+2)}-1 \right].
\end{align}
Here we have used
\begin{equation}
    \psi(z+1)-\psi(z)=\frac{1}{z}
\end{equation}
to take all the polygamma functions on $n$ into $\psi(n+1)$. The divergent part is contributed by the $-1$ in the parenthesis. We can regularize it by taking derivatives on $T$ and integrating it back, similar as described in \cref{sec:resum}, which leads to
\begin{equation}
    \sum_{n=0}^\infty -\frac{1}{T-n} = -\psi(-T) + C.
\end{equation}
The rest of the sum can be worked out by
\begin{align}
    \sum_{n=0}^\infty \frac{1}{(n+b)(T-n)} =& \frac{1}{T+b}\big[\psi(-T)-\psi(b)\big],\\
    \sum_{n=0}^\infty \frac{\psi(n+1)}{(n+b)(T-n)} =& \frac{1}{2(T+b)}\left[\psi(-T)^2-\psi(b)^2 - \psi'(-T) + \psi'(b) \right].
\end{align}
One can verify that under certain simplifications this gives the same result as \eqref{eq:gluonresum2222}.

\section{The seed functions in terms of MPLs}\label{appx:function}

In this appendix, we intend to preform some complete expressions of some seed functions as a supplement to the main content. Let us give some explicit expression of our seed function in MPLs,
\begin{align}
	W_{2}(z,\zb)=&\ G_{0,\zb}G_{1,z}+G_{01,\zb}+G_{10,z}-(z\leftrightarrow \zb)\,, \\
	W_{3}(z,\zb)=&\  G_{00,\zb} G_{1,z}-G_{1,\zb} G_{00,z}+G_{01,\zb} G_{0,z}-2 G_{01,\zb} G_{\zb ,z}+G_{0,\zb} G_{01,z} -G_{10,\zb} G_{0,z}\nonumber \\
	&-G_{10,\zb} G_{1,z}+2 G_{10,\zb} G_{\zb ,z}+G_{0,\zb} G_{10,z}-G_{1,\zb} G_{10,z} +2 G_{1,\zb} G_{\zb 0,z}\nonumber  \\
	&-2 G_{0,\zb} G_{\zb 1,z}+2 G_{\zb 01,z}-2 G_{\zb 10,z}-G_{001,\zb}+G_{010,\zb} -G_{100,\zb}\nonumber\\
	&+G_{101,\zb}-G_{001,z}+G_{010,z}+G_{100,z}-G_{101,z}\, , \\
	W_{4}(z,\zb)=&\ G_{01,\zb} G_{00,z}+G_{10,\zb} G_{01,z}+G_{11,\zb} G_{10,z}+G_{00,\zb} G_{11,z} +G_{001,\zb} G_{1,z}\nonumber \\
	&+G_{0,\zb} G_{001,z}+G_{010,\zb} G_{0,z}+G_{1,\zb} G_{010,z}  +G_{101,\zb} G_{0,z}+G_{1,\zb} G_{101,z}\nonumber \\ 
	&+G_{110,\zb} G_{1,z } +G_{0,\zb } G_{110,z}+G_{0011,\zb}+G_{0100,\zb}+G_{1010,\zb}+G_{1101,\zb} \nonumber \\ 
	&+G_{0010,z}+G_{0101,z}+G_{1011,z}+G_{1100,z} - (z\leftrightarrow \zb) \, .
\end{align}
The expressions of other seed functions are too heavy to fit them here so we provide them in the supplement material of this paper.

\bibliography{refs}
\bibliographystyle{utphys}

\end{document}